\documentclass[pra,aps,amsmath,amssymb,amsfonts,twocolumn,nofootinbib,longbibliography,floatfix,superscriptaddress]{revtex4-1}
\usepackage{amssymb}
\usepackage{bm,mathrsfs}
\usepackage{graphicx}
\usepackage{epsfig}
\usepackage{amsmath,bbm}
\usepackage{amsfonts,amssymb}
\usepackage{times}
\usepackage{verbatim}
\usepackage[sort&compress]{natbib}
\usepackage{amsmath}
\usepackage{bm}
\usepackage{lipsum}

\usepackage{makecell}
\usepackage{array}
\usepackage{booktabs}
\usepackage{multirow}
\usepackage[colorlinks,breaklinks,linkcolor=blue,anchorcolor=blue,citecolor=blue,urlcolor=blue]{hyperref}
\begin{document}
\title {Reservoir-assisted quantum battery charging at finite temperatures}
\author{Y. Yao}
\affiliation{School of Science, Shenyang Ligong University, Shenyang, Liaoning 110159, China}

\author{X. Q. Shao}
\email{shaoxq644@nenu.edu.cn}
\affiliation{Center for Quantum Sciences and School of Physics, Northeast Normal University, Changchun, Jilin 130024, China}
\affiliation{Center for Advanced Optoelectronic Functional Materials Research and Key Laboratory for UV Light-Emitting Materials and Technology, Ministry of Education, Northeast Normal University, Changchun 130024, China}

\begin{abstract}
Quantum batteries, as highly efficient energy storage devices, have garnered significant research interest. A key challenge in their development is to maximize the extractable energy (ergotropy) when operating within a finite-temperature reservoir. To address this, we apply quantum feedback control to the charger and investigate the effects of fermionic and bosonic thermal reservoirs on the performance of quantum batteries, including stored energy, ergotropy, and charging efficiency, in an open environment.
Our findings reveal that, regardless of the type of thermal reservoir, the system exhibits optimal charging parameters. In particular, in a fermionic thermal reservoir, increasing the environmental temperature enhances battery performance, enabling stable and efficient charging. In contrast, within a bosonic thermal reservoir, higher temperatures hinder energy storage and extraction, significantly reducing charging efficiency. Additionally, we explore the impact of battery size and find that, under a fermionic reservoir, increasing the battery size appropriately can further improve performance.
 \end{abstract}

\maketitle
\section{Introduction}
The growing demand for energy resources and the increasing environmental challenges have sparked a growing interest in the investigation of revolutionary energy storage and supply devices. Researchers anticipate that by integrating experimental techniques for precise detection and manipulation at the qubit level, they can realize a new principle of the energy storage and supply device: the quantum battery \cite{2013}. Unlike traditional batteries, which convert chemical energy or other forms of energy storage into electricity, quantum batteries store energy based on the principles of quantum mechanics, typically in the form of excited states of quantum systems. At present, quantum batteries have been explored in a variety of physical systems, including atoms and molecules \cite{atom1,atom2,atom3,atom4,fen1}, spins~\cite{spin1,v12,chen2022quantum,v15,Yang_2025,v7,chain2,chain3,chain4,chain5,chain6}, transmon \cite{dou2023superconducting,gemme2024qutrit,yang2024resonator}, micromaser \cite{micromasers1,micromasers3,micromasers4}, etc. They are expected to be smaller, possess a higher charging power \cite{v1,v2,v3,v4,v5,v6,v13,v8,v9,v10,v11,v17,v18,v14,PhysRevLett.134.130401} and higher charging capacity \cite{chu1,chu2,chu6,chu3,chu7,santos2023vacuum,chu4,chu5,wang2025dynamics,cavaliere2025dynamical}, and offer a greater amount of extractable work \cite{chuTi5,chuTi1,chuTi7,chuTi2,chuTi3,chuTi4,chuTi6,bhattacharyya2024,elghaayda2025performance} compared to traditional batteries.

Open quantum batteries, which are closer to practical application scenarios, have greater research value than ideal closed quantum batteries, but they face several challenges. For example, environmentally induced decoherence causes the stored energy in quantum batteries to spontaneously dissipate, leading to battery aging and reduced charging efficiency \cite{laohua1}. Additionally, a finite-temperature environment will cause the battery to tend toward a passive state, thereby hindering the extraction of energy from the quantum battery \cite{r1,r2,r3}. In short, as a cutting-edge technology, the quantum battery has made progress in theory and experiment \cite{shiyan1,IBM,fen2,Hu_2022,shiyan3,shiyan4,Colloquium,Razzoli_2025,lu2025topological}, but to realize its commercialization and large-scale application, many technical challenges remain to be overcome, including environmental decoherence and energy extraction efficiency.

Considering the potential impact of environmental factors on the charging performance of the battery, researchers have designed a variety of charging schemes. These include using adiabatic \cite{juere1,juere4,juere2,juere3,Hu_2022}, measurement \cite{m1,m2,m4,m3,chaki2024}, and control techniques \cite{fed1,yao,yao2,control2,control1} to develop the corresponding strategies that ensure open quantum batteries can be charged stably and efficiently. Several schemes utilizing Floquet engineering \cite{laohua2} and environment engineering \cite{en1,en1a,en2,en4,en8,en3,en6,en9,en5,en7,ahmadi2025superoptimal} aim to beat the decoherence effects. In addition, some researchers have also paid attention to the influence of temperature on the performance of quantum batteries. For example, Song {\it et al}. \cite{Song_2024} studied the extractable work and conversion efficiency of a quantum battery in the presence of a bosonic or fermionic thermal reservoir using the entropy uncertainty relation. Under the positive temperature mechanism, the temperature reduces the energy extraction and conversion efficiency of the battery. A feature of the aforementioned studies is that they treat the battery as an ideal non-dissipative model, whereas the dissipation of the battery itself should be accounted for in practical scenarios. Kamin {\it et al}. \cite{kamin2024} studied the steady-state charging process of a single-cell quantum battery embedded in the center of a star comprising $N$ qubits, under two different equilibrium and nonequilibrium scenarios. The results indicate that high temperature has a detrimental effect across all parameter ranges, inhibiting the battery's energy extraction. Quach and Munro\cite{dark1} used dark states to design quantum batteries with both superextensive capacity and power density. However, an increase in temperature is detrimental to energy storage in the quantum battery and reduces the upper limit of the battery's energy extraction. Cruz {\it et al}. \cite{cruz} put forward a feasible realization of a quantum battery based on carboxylate metal complexes. The scheme demonstrates that temperature hinders battery energy extraction, with the extracted energy being 75$\%$ of its corresponding maximum at room temperature. Therefore, improving the maximum extractable energy and charging efficiency of the battery in finite-temperature reservoirs remains a challenge to address.

In this work, our objective is to design an open quantum battery charging scheme that ensures stable battery charging while transforming finite temperatures into a favorable charging factor. The battery charging process is assisted by a charger which operates under homodyne quantum feedback control. This real-time feedback enables precise control of the quantum system, allowing more energy to be transferred into the battery. On the basis of this approach, we investigate the impact of bosonic and fermionic reservoirs with finite temperatures on the performance of quantum batteries. Unlike bosonic thermal reservoirs, fermionic reservoirs are typically modeled as two-level systems \cite{fermi1,fermi4,fermi3,fermi2,fermi5,fermi6}, with their components having discrete energy levels. Some studies are also exploring the potential advantages of using this type of reservoir \cite{kamin2024,Song_2024,fery6,fery4, fery3, fery2, fery1,fery5}.

In our proposal, we consider both the charger and the quantum battery to be modeled as two-level systems, each embedded in its respective environment. The remainder of the paper is organized as follows. In Sec.~\ref{II}, the charging model of the quantum battery is introduced and related physical quantities characterizing the performance of the quantum battery are provided, such as energy storage, maximum extractable energy, and charging efficiency of the battery. In Sec.~\ref{III}, the charging process of the quantum battery in bosonic and fermionic reservoirs is discussed. The results show that there exist optimal charging parameters, and a fermionic reservoir at a finite temperature is more conducive to enhancing the battery's performance. In Sec.~\ref{IV}, the single-particle quantum battery model is extended to a multiparticle quantum battery, and the impact of increasing the number of particles in the quantum battery on its performance is discussed. We give a summary of the present protocol in Sec.~\ref{V}.

\section{DISSIPATIVE QUANTUM BATTERY MODEL}\label{II}
\subsection{Model and the master equation of the system}
\begin{figure}
\centering\scalebox{0.4}{\includegraphics{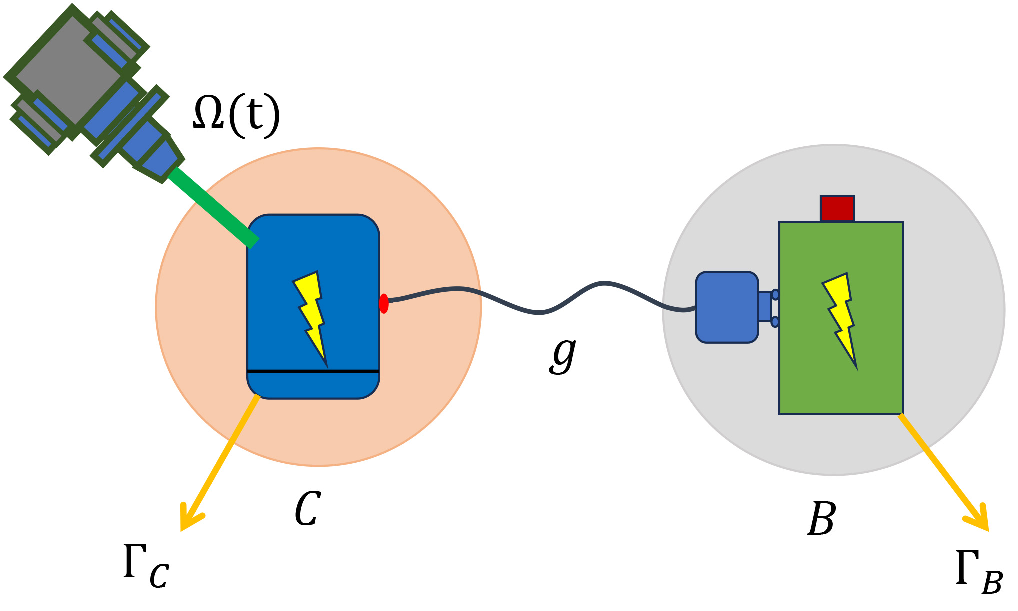}}
\caption{\label{Q1} Quantum battery model. Both the charger $C$ and the quantum battery $B$ are modeled as two-level systems and exposed to their respective dissipative environments. The corresponding transition frequencies are $\omega_{C}$ and $\omega_{B}$, respectively. The direct coupling strength between the charger and the quantum battery is $g$. The charger is driven by an external field and subject to homodyne feedback control.
}
\end{figure}

We model the charger and the quantum battery as a two-level atom with an energy interval of $\omega_{C}$ and $\omega_{B}$ between the excited and ground states, respectively. Meanwhile, the charger and quantum battery are injected into their respective reservoirs, as shown in Fig.~\ref{Q1}. Initially, the quantum battery is in the ground state $|g\rangle$. For convenience, we choose $\hbar=1$ throughout the paper, and we assume that the charger and quantum battery share the same transition frequency, that is, $\omega_{C} = \omega_{B} =\omega_{0}$. The total Hamiltonian of the system is $H = H_{0}+ H_{I}$, where
\begin{equation}\label{H0}
H_{0}=\omega_{0}\sigma_{C}^{+}\sigma_{C}^{-}+\omega_{0}\sigma_{B}^{+}\sigma_{B}^{-}+\sum\limits_{k}\omega_{Ck}a^{\dag}_{k}a_{k}+\sum\limits_{k}\omega_{Bk}b^{\dag}_{k}b_{k}.
\end{equation}
The first two terms represent the free Hamiltonian of the charger and the quantum battery, while the last two terms correspond to the free Hamiltonian of the independent environments interacting with the charger and the battery, respectively. In addition, $\sigma_{j}^{+}$ and $\sigma_{j}^{-}$ $(j=C, B)$ are the raising and lowering operators for the corresponding systems, respectively, and $a^{\dag}_{k}$ $(a_{k})$ and $b^{\dag}_{k}$ $(b_{k})$, and $\omega_{jk}$ $(j=C, B)$ are the creation (annihilation) operators and the frequency of the $k$th mode of the corresponding environments. In the interaction picture, the interacting Hamiltonian involved in the system is $H_{ I}=H_{SI}+H_{EI}$, with
\begin{eqnarray}\label{HI}
H_{SI}&=&g(\sigma_{C}^{+}\sigma_{B}^{-}+\sigma_{C}^{-}\sigma_{B}^{+}),
\end{eqnarray}
\begin{eqnarray}\label{HI}
H_{EI}&=&\sum\limits_{k}g_{Ck}(\sigma_{C}^{+}a_{k}e^{i(\omega_{0}-\omega_{k})t}+\sigma_{C}^{-}a_{k}^{\dag}e^{-i(\omega_{0}-\omega_{k})t})\nonumber\\&&+\sum\limits_{k}g_{Bk}(\sigma_{B}^{+}b_{k}e^{i(\omega_{0}-\omega_{k})t}+\sigma_{B}^{-}b_{k}^{\dag}e^{-i(\omega_{0}-\omega_{k})t}),
\end{eqnarray}
where $g$ stands for the coupling strength between the charger and the quantum battery and $g_{jk}$ $(j=C, B)$ is the coupling strength of the system with the respective $k$th mode environment. For simplicity, we assume $\omega_{Ck}=\omega_{Bk}=\omega_{k}$. During the charging process, a resonant drive is applied to the charger and then the charger resonates with the battery to exchange energy through the interaction. In the interaction picture and considering the rotating-wave approximation, the corresponding driving Hamiltonian is
\begin{equation}\label{Hdrive}
H_{\rm drive}=\Omega\sigma_{C}^{y},
\end{equation}
where $\Omega$ is the strength of the drive field, which is usually constant in conventional schemes.

To improve the charging efficiency of the battery, a feedback control is applied to the charger. The photons spontaneously emitted by the charger are collected and measured by a homodyne interferometer, thus obtaining the corresponding photocurrent represented by an appropriately normalized and shifted measurement record \cite{hom1,hom2,carmichael2009open}
\begin{equation}\label{J}
r(t)dt=\langle\sigma_{C}^{x}\rangle dt+\frac{dw(t)}{\sqrt{\eta\Gamma_{C}}},
\end{equation}
where $\langle\sigma_{C}^{x}\rangle$ stands for the corresponding expected value; $dw(t)$ represents the Wiener increment, which satisfies $[dw(t)]^{2}=dt$ and $E[dw(t)]=0$; $\Gamma_{C}$ is the spontaneous emission rate of the charger; and $\eta$ is the total measurement efficiency, which incorporates both the fraction of collected photons $\eta_{c}$ and the detector efficiency $\eta_{d}$ so that $\eta=\eta_{c}\eta_{d}$. The feedback is achieved by applying a driving field that depends on the measured results. We consider the simplest case of direct feedback, where the drive strength is proportional to the measurement record, i.e.,
\begin{equation}\label{JJ}
\Omega(t)=\Omega_{0}-fr(t-\tau)dt,
\end{equation}
where $\Omega_{0}$ is a constant drive, which is set to zero here; $f$ stands for the feedback strength; and $\tau$ expresses a small time delay in the feedback loop. Here we consider the measurement efficiency to be $\eta=1$ and assume that the decay channel corresponding to photons collected by the detector remains at effectively zero temperature. Simultaneously, the quantum battery is considered to be in contact with a thermal reservoir at a finite temperature. In the Markovian limit of $\tau\rightarrow0$ and weak-coupling conditions \cite{Hofer_2017,Potts_2021,PhysRevResearch.6.043091}, the evolution of the system based on feedback control can be determined by the master equation \cite{weiner2}
\begin{eqnarray}\label{master}
\mathop{\dot{\rho}}&=&-i[H_{\rm SI},\rho]+if[\sigma_{C}^{y},\sigma_{C}^{-}\rho+\rho\sigma_{C}^{+}]+\frac{f^{2}}{\eta\Gamma_{C}}\mathcal{D}[\sigma_{C}^{y}]\rho\nonumber\\&&+\mathcal{L}_{C}\rho+\mathcal{L}_{B}\rho.
\end{eqnarray}
The first term of Eq.~(\ref{master}) describes the coherent evolution introduced by the interaction between the charger and the quantum battery. The last two terms in the first line describe the coherent evolution introduced by the feedback operation to the charger, and the measurement noise fed back to the system by the driver. The two terms in the second line describe the charger's spontaneous emission and the quantum battery's dissipation process, respectively.

The specific forms of $\mathcal{L}_{C}\rho$ and $\mathcal{L}_{B}\rho$ are as follows:
$\mathcal{L}_{C}\rho=\Gamma_{C}[\sigma_{C}^{-}\rho\sigma_{C}^{+}-1/2(\sigma_{C}^{+}\sigma_{C}^{-}\rho+\rho\sigma_{C}^{+}\sigma_{C}^{-})]$ and $\mathcal{L}_{B}\rho=\gamma^{\downarrow}_{B}[\sigma_{B}^{-}\rho\sigma_{B}^{+}-1/2(\sigma_{B}^{+}\sigma_{B}^{-}\rho+\rho\sigma_{B}^{+}\sigma_{B}^{-})]+\gamma^{\uparrow}_{B}[\sigma_{B}^{+}\rho\sigma_{B}^{-}-1/2(\sigma_{B}^{-}\sigma_{B}^{+}\rho+\rho\sigma_{B}^{-}\sigma_{B}^{+})]$. Here we consider the case where the dynamics of the quantum battery is controlled by bosonic or fermionic thermal reservoirs, respectively. If the quantum battery interacts with a bosonic thermal reservoir, $\gamma^{\downarrow}_{B}=\Gamma_{B}(1+n_{b})$ and $\gamma^{\uparrow}_{B}=\Gamma_{B}n_{b}$, where $\Gamma_{B}$ is the dissipation rate of the quantum battery and $n_{b}=1/(e^{1/T}-1)$ is the average excitation number of the corresponding thermal reservoir, in which $T$ is the dimensionless temperature of the thermal reservoir. If the quantum battery interacts with a fermionic thermal reservoir, $\gamma^{\downarrow}_{B}=\Gamma_{B}(1-n_{f})$, $\gamma^{\uparrow}_{B}=\Gamma_{B}n_{f}$, and $n_{f}=1/(e^{1/T}+1)$. In the case of the fermionic thermal reservoir, it is easy to see that $0<n_{f}<1$, where $0<n_{f}<1/2$ ($1/2<n_{f}<1$) means $T>0$ ($T<0$). Meanwhile, a negative temperature $T<0$ can exhibit stationary states with population inversion \cite{negative3,negative4,negative6,negative1,negative2,negative7,negative5,negative8,negative9}. Although the temperature of the fermionic thermal reservoir can be positive or negative, it is worth noting that the bosonic reservoir cannot attain negative temperatures, resulting in $0<n_{b}<\infty$. In this work, we limit the temperature of the fermionic thermal reservoir to the positive-temperature range ($T>0$) to avoid population inversion induced by the thermal reservoir effects and then compare the effects of the fermionic thermal reservoir and the bosonic thermal reservoir on battery performance.
\subsection{Related performance parameter of the quantum battery}
At an arbitrary time $t$, the stored energy of the quantum battery is defined as
\begin{equation}\label{energy}
 E_{B}(t)={\rm Tr}[H_{B}\rho_{B}(t)]-{\rm Tr}[H_{B}\rho_{B}(0)],
\end{equation}
where $H_{B}=\omega_{0}\sigma_{B}^{+}\sigma_{B}^{-}$, $\rho_{B}(t)={\rm Tr}_{C}[\rho(t)]$, and $\rho_{B}(0)=|g\rangle_{B}\langle g|$ are the free Hamiltonian, the reduced density matrix, and the initial state of the quantum battery, respectively. When the quantum battery is in the excited state $|e\rangle$, it means that it is fully charged, and the corresponding maximum stored energy is $E_{B}^{\rm max}=\omega_{0}$. To evaluate the useful work of the quantum battery storage, we use the concept of ergotropy \cite{ergotropy,ergotropy2} and its form is
\begin{equation}\label{ergotropy}
\mathcal{E}(t)={\rm Tr}[H_{B}\rho_{B}(t)]-\mathop{\rm min}\limits_{U}{\rm Tr}[U\rho_{B}(t)U^{\dag}H_{B}],
\end{equation}
which describes the maximum energy that can be extracted from the battery. The second term in Eq.~(\ref{ergotropy}) indicates the minimum energy that cannot be extracted from the battery by executing all unitaries $U$ in the system, and the state corresponding to the minimum energy is known as the passive state \cite{Passive,Passive2}.

The Hamiltonian $H_{B}$ and the density matrix $\rho_{B}$ of the quantum battery can be written in order using spectral decomposition as $H_{B}=\sum_{j}\varepsilon_{j}|\varepsilon_{j}\rangle\langle \varepsilon_{j}|$ ($\varepsilon_{1}\leq\varepsilon_{2}\cdots\leq\varepsilon_{n}$), and $\rho_{B}=\sum_{j}r_{j}|r_{j}\rangle\langle r_{j}|$ ($r_{1}\geq r_{2}\cdots\geq r_{n}$), respectively, the form of the passive state can be expressed as $\sigma=\sum_{j}r_{j}|\varepsilon_{j}\rangle\langle \varepsilon_{j}|$. Thus, the ergotropy can be rewritten as
\begin{eqnarray}\label{ergotropy1}
\mathcal{E}(t)&=&{\rm Tr}[H_{B}\rho_{B}(t)]-{\rm Tr}(H_{B}\sigma)\nonumber\\
         &=&{\rm Tr}[H_{B}\rho_{B}(t)]-\sum_{j}r_{j}\varepsilon_{j}.
\end{eqnarray}
In addition to the stored energy and extractable energy of the quantum battery, the charging efficiency at steady state is another crucial performance metric. It is defined as the ratio of the maximum extractable energy to the stored energy, i.e.,
\begin{equation}\label{R}
R=\frac{\mathcal{E}(\infty)}{E_{B}(\infty)}.
\end{equation}
The higher this ratio, the more useful work the battery can store, and the better its performance
\section{CHARGING PROCESS OF A QUANTUM BATTERY IN DIFFERENT THERMAL RESERVOIR ENVIRONMENTS}\label{III}
\subsection{Charging process of quantum battery in a bosonic thermal reservoir}
In this section, we first consider the case where the quantum battery is in a bosonic thermal reservoir. After setting $\mathop{\dot{\rho}}=0$ in Eq.~(\ref{master}), the steady-state solutions of the system are obtained (see Appendix \ref{A}). Combined with Eq.~(\ref{energy}), the stored energy of the battery is finally obtained as 
\begin{equation}\label{E12}
E_{B}{(\infty)}=\omega_{0}\frac{n_{b}W_{b}S+4g^{2}Q_{b}[2Q_{b}+\Gamma_{C}(1-2\delta)\eta+\Gamma_{B}\eta]}{(1+2n_{b})W_{b}S+4g^{2}[2Q_{b}+\Gamma_{C}(1-2\delta)\eta+\Gamma_{B}\eta]^{2}},
\end{equation}
where we set $\delta=f/\Gamma_{C}$, $S=2\delta(\delta-\eta)+\eta$, $Q_{b}=\Gamma_{C}\delta^{2}+n_{b}\Gamma_{B}\eta$ and $W_{b}=\Gamma_{C}\Gamma_{B}(\Gamma_{C}+\Gamma_{B}+2n_{b}\Gamma_{B})[4Q_{b}+(\Gamma_{C}+\Gamma_{B}-4\Gamma_{C}\delta-2n_{b}\Gamma_{B})\eta]$ for simplicity. See Appendix \ref{A} for the corresponding calculation details. Under perfect measurement conditions $(\eta=1)$, we assume that the quantum battery is in an ideal closed state, which means that $\Gamma_{B}=0$. At this time, the energy stored in the battery becomes
\begin{equation}
E_{B}(\infty)=\omega_{0}\frac{\delta^{2}}{1+2(\delta-1)\delta}.
\end{equation}
This clearly shows that when $\delta=f/\Gamma_{C}=1$, the stored energy of the battery reaches its maximum value, that is, $E_{B}^{\rm max}(\infty)=\omega_{0}$. This implies that the effect of spontaneous emission on the charger can be offset by adjusting the feedback intensity $f$, enabling full charging of the quantum battery, this result is consistent with Ref.~\cite{fed1}. However, in practice, a quantum battery cannot be perfectly isolated from its surroundings, so it is necessary to consider placing it in an open system. To discuss the influence of feedback strength $f$ and charger dissipation on quantum battery energy storage, we conduct corresponding numerical simulations according to Eq.~(\ref{E12}). In the absence of special instructions, the remaining parameters are set as $\eta=1$, $g=0.01\omega_{0}$, and $\Gamma_{B}=0.1g$.
\begin{figure}
\centering\scalebox{0.27}{\includegraphics{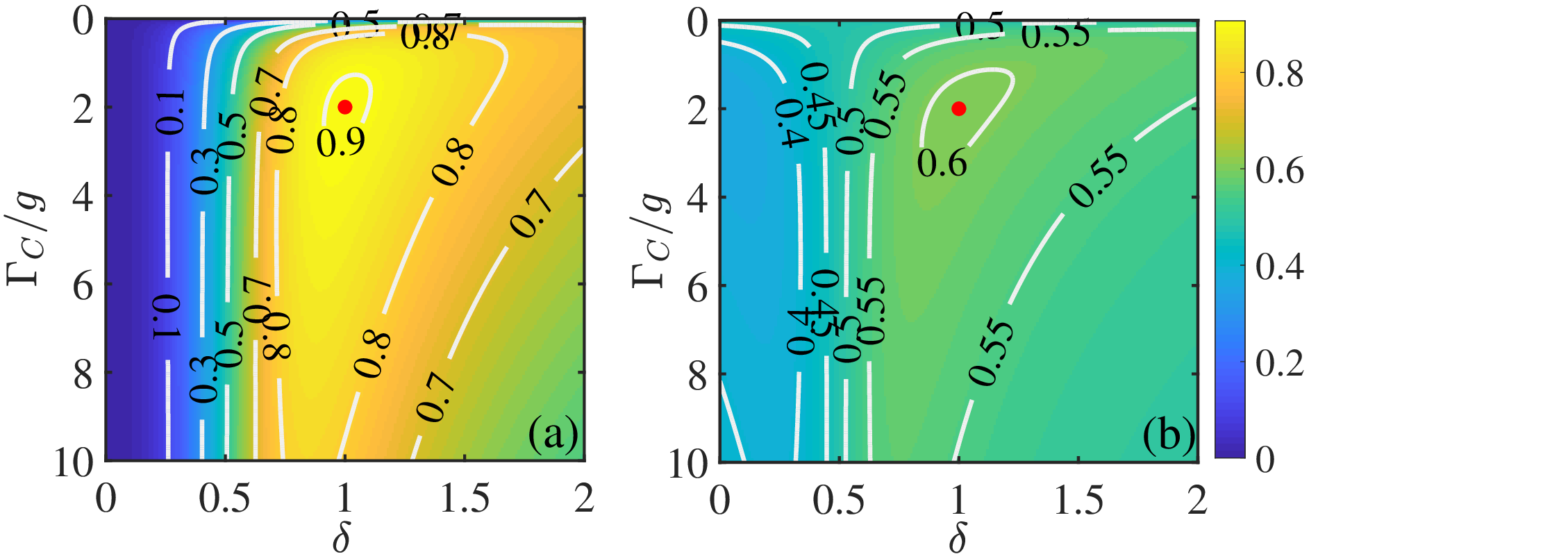}}
\caption{\label{Q2} Under a bosonic thermal reservoir, the stored energy of the quantum battery is a function of the feedback parameter $\delta$ and the dissipation rate $\Gamma_{C}$ of the charger. The temperature of the bosonic thermal reservoir is (a) $T=0$ $(n_{b}=0)$ and (b)$T=10$ ($n_{b}\approx9.51$). Red circles represent the optimal stored energy of the quantum battery, with the corresponding optimal charging parameters being $\Gamma_{C}=2g$ and $\delta=f/\Gamma_{C}=1$. The other parameters are $g=0.01\omega_{0}$ and $\Gamma_{B}=0.1g$, respectively.
}
\end{figure}

Figures~\ref{Q2}(a) and \ref{Q2}(b) show the stored energy of the quantum battery as a function of the feedback parameter $\delta$ and the dissipation rate $\Gamma_{C}$ of the charger when the battery is in a dissipative bosonic thermal reservoir with temperatures $T=0$ ($n_{b}=0$) and $T=10$ ($n_{b}\approx9.51$), respectively. The results show that the optimal stored energy of the battery is regulated by the feedback parameter $\delta$ and the dissipation rate $\Gamma_{C}$ of the charger. There exists a set of optimal parameters $\Gamma_{C}=2g$ and $\delta=f/\Gamma_{C}=1$. Here $\delta=1$ means that the feedback strength $f$ is equal to the dissipation rate $\Gamma_{C}$ of the charger (i.e.,$f=\Gamma_{C}$).

\begin{figure}
\centering\scalebox{0.28}{\includegraphics{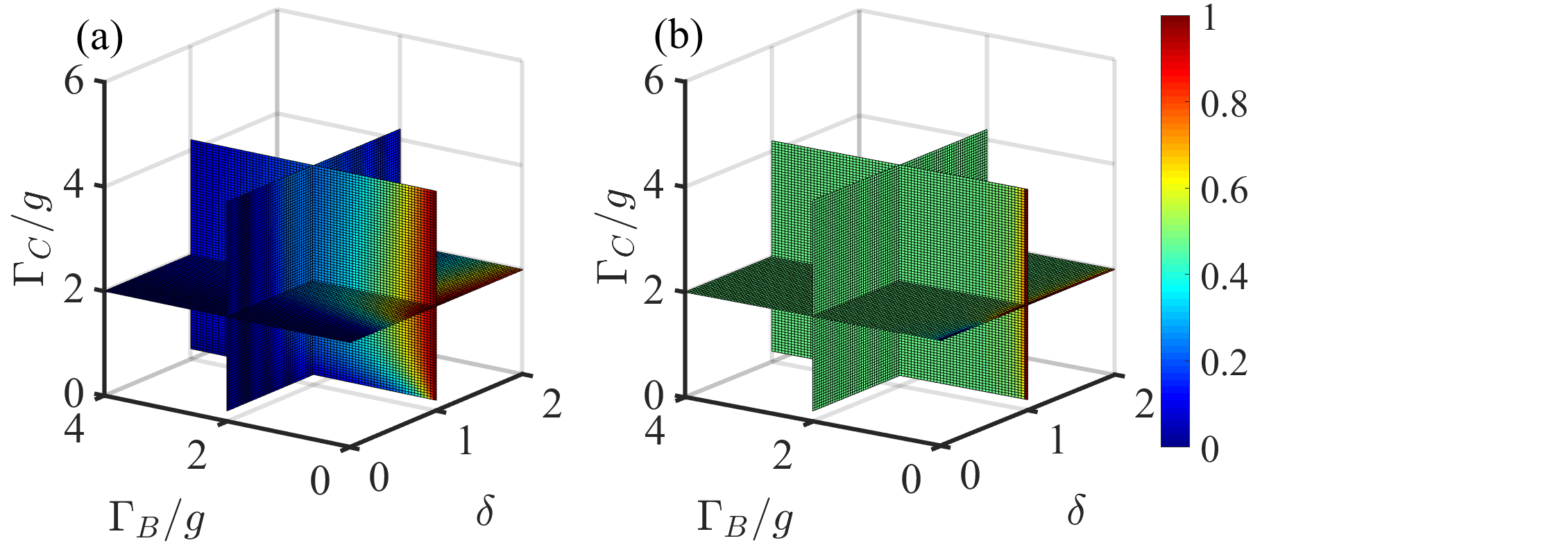}}
\caption{\label{Q3} Steady-state stored energy of the quantum battery plotted as a function of $\delta$, $\Gamma_{B}$, and $\Gamma_{C}$. The temperature of the thermal reservoir is set to (a) $T=0$ and (b) $T=10$. The remaining parameters are the same as in Fig.~\ref{Q2}.
 }
\end{figure}

To determine whether optimal charging conditions are affected by the dissipation rate of the battery $\Gamma_{B}$, we explore the influence of the feedback parameter $\delta$ and the dissipation rate of the charger and the quantum battery on the stored energy of the quantum battery in steady state in Fig.~\ref{Q3}. The thermal reservoir temperatures for Figs.~\ref{Q3}(a) and \ref{Q3}(b) are $T=0$ ($n_{b}=0$) and $T=10$ ($n_{b}\approx9.51$), respectively. The results show that when the battery dissipation rate is zero ($\Gamma_{B}=0$), there is no restriction on the magnitude of the dissipation rate $\Gamma_{C}$ of the charger; as long as the optimal feedback parameter satisfies $\delta=f/\Gamma_{C}=1$, the quantum battery can be fully charged. This is consistent with the results of the previous analysis. However, when the quantum battery is coupled to a dissipative environment, it can be observed that, under different battery dissipation rates, the stored energy of the quantum battery reaches its optimal value only when $\delta=1$ and $\Gamma_{C}=2g$. This demonstrates that the nonzero dissipation rate of the quantum battery does not affect its optimal charging conditions. In addition, when the temperature of the thermal reservoir and the dissipation rate of the battery are high, as shown in Fig.~\ref{Q3}(b), the stored energy of the battery does not change significantly with its dissipation rate.
\begin{figure}
\centering\scalebox{0.36}{\includegraphics{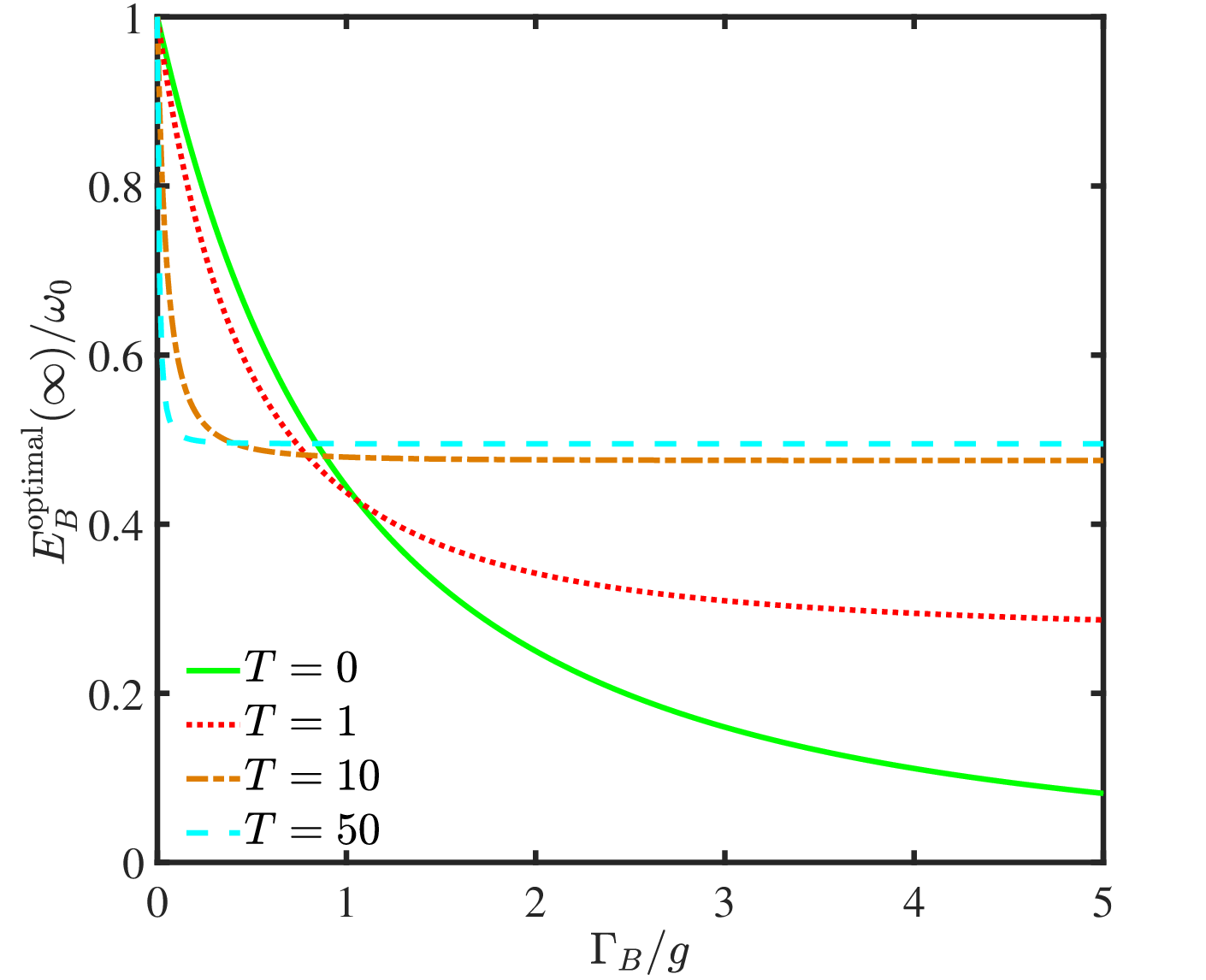}}
\caption{\label{Q4} Under optimal charging conditions $\Gamma_{C}=2g$ and $\delta=1$, the stored energy of the quantum battery in steady state varies with the battery dissipation rate. Different curves represent different temperatures of the bosonic thermal reservoir. The green solid line, red dotted line, yellow dash-dotted line, and blue dashed line correspond to the bosonic thermal reservoir temperatures $T$ of 0, 1, 10, and 50, respectively. The average thermal photon number as a function of temperature is given by $n_{b}=1/(e^{1/T}-1)$, where $T=0$ corresponds to $n_{b}=0$. All other parameters are the same as in Fig.~\ref{Q2}.
 }
\end{figure}

Under the perfect measurement ($\eta=1$) and the optimal charging conditions ($\Gamma_{C}=2g$ and $\delta=1$), the stored energy of the battery becomes
\begin{equation}\label{E14}
E_{B}^{\rm optimal}(\infty)=\omega_{0}\frac{4g^{2}+4gn_{b}\Gamma_{B}+n_{b}(1+2n_{b})\Gamma_{B}^{2}}{(2g+\Gamma_{B}+2n_{b}\Gamma_{B})^{2}}.
\end{equation}
Based on Eq.~(\ref{E14}), we explore the influence of quantum battery dissipation $\Gamma_{B}$ on optimal steady-state energy storage at different temperatures in Fig.~\ref{Q4}, where each curve corresponds to a different temperature. It can be seen that when the dissipation rate $\Gamma_{B}$ of the battery is low, the optimal steady-state stored energy of the battery decreases with increasing temperature. When the dissipation rate $\Gamma_{B}$ of the battery gradually increases, the optimal stored energy of the battery increases and approaches a constant value as the temperature increases. It is easy to see from Eq.~(\ref{E14}) that when $\Gamma_{B} \gg g$ and $n_{b}\rightarrow \infty$ the battery stored energy $E_{B}^{\rm optimal}(\infty)\approx 0.5\omega_{0}$.

\begin{figure}
\centering\scalebox{0.27}{\includegraphics{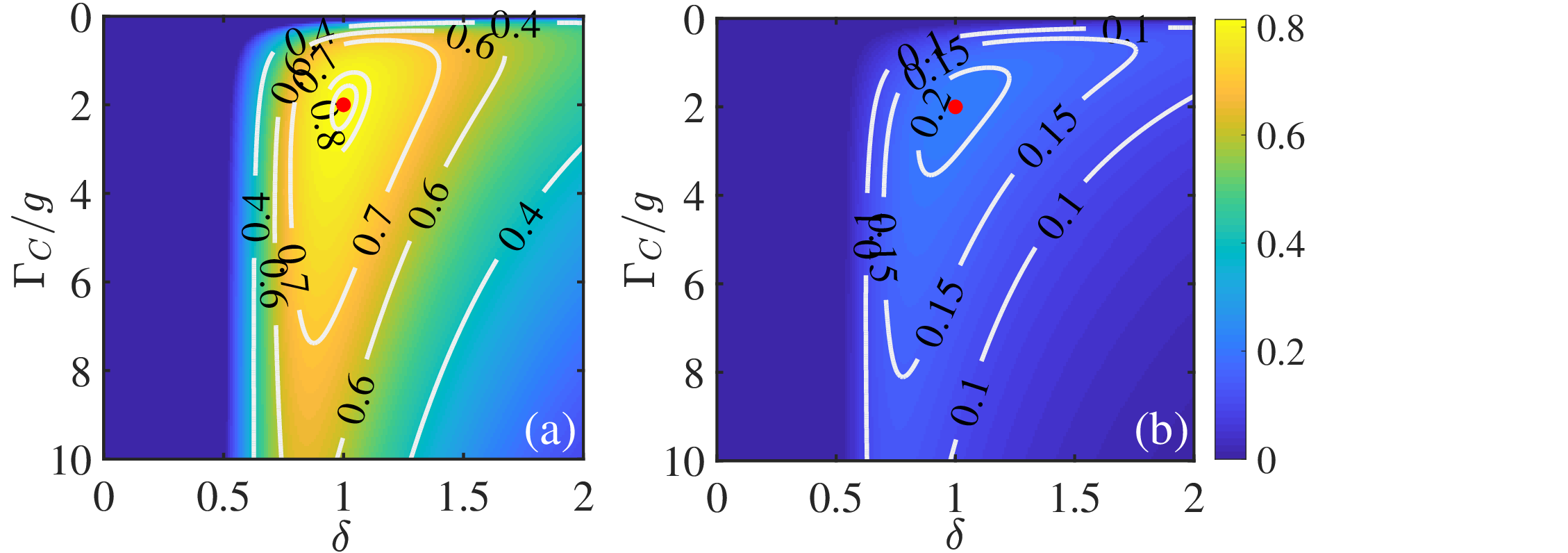}}
\caption{\label{Q5} The quantum battery is placed in a bosonic thermal reservoir. The ergotropy of the quantum battery is depicted as a function of the feedback parameter $\delta$ and the dissipation rate $\Gamma_{C}$ of the charger in the steady state for environmental temperatures corresponding to (a) $T=0$ and (b) $T=10$. Red circles represent the optimal ergotropy of the battery, with the corresponding optimal charging parameters being $\Gamma_{C}=2g$ and $\delta=f/\Gamma_{C}=1$. The other parameters are the same as in Fig.~\ref{Q2}.
 }
\end{figure}
\begin{figure}
\centering\scalebox{0.28}{\includegraphics{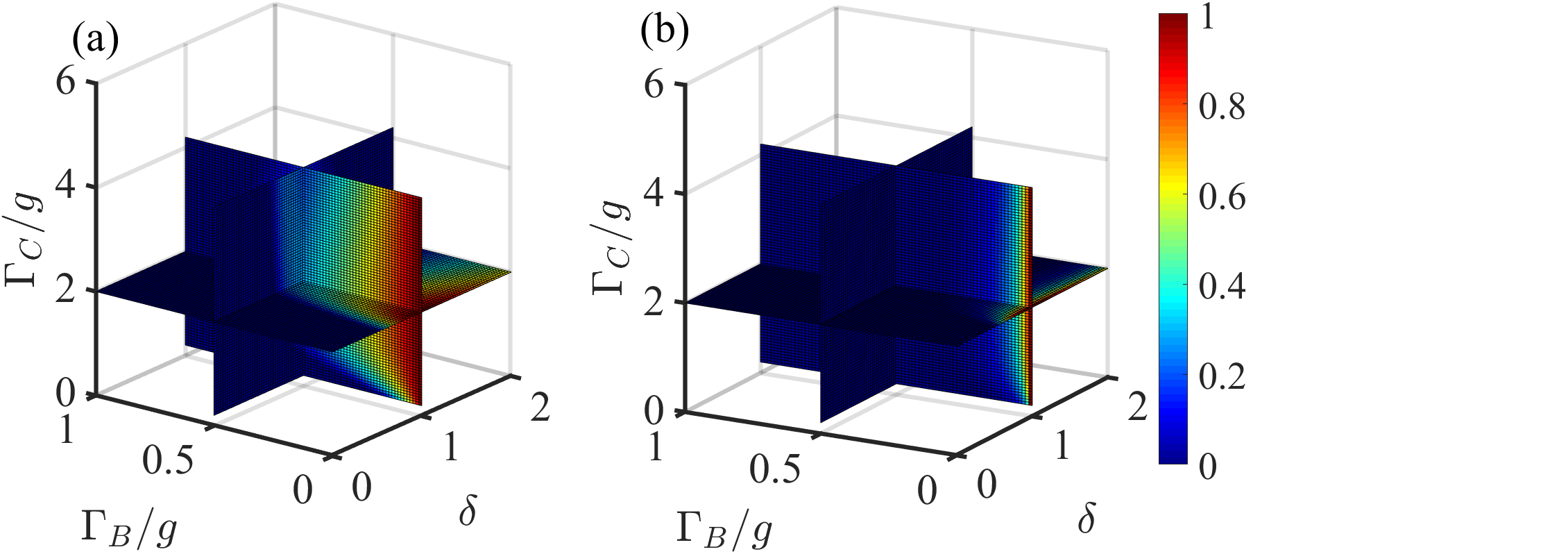}}
\caption{\label{Q6} Steady-state ergotropy of the quantum battery as a function of $\delta$, $\Gamma_{B}$, and $\Gamma_{C}$, with the thermal reservoir temperatures set to (a) $T=0$ and (b) $T=10$. The remaining parameters are identical to those in Fig.~\ref{Q2}.
 }
\end{figure}

In addition to focusing on the stored energy of the quantum battery, the ergotropy of the quantum battery is also an important performance evaluation indicator. Similarly to Fig.~\ref{Q2}, Fig.~\ref{Q5} plots the ergotropy of the quantum battery in the steady state as a function of the feedback parameter $\delta$ and the dissipation rate $\Gamma_{C}$ of the charger, where the dissipation rate of the battery is $\Gamma_{B}=0.1g$. The results demonstrate that optimal charging parameters still exist for the battery ergotropy. These optimal conditions coincide with those for battery energy storage, namely, $\Gamma_{C}=2g$ and $\delta=1$. As shown in Fig.~\ref{Q6}, the optimal charging conditions for the ergotropy of the quantum battery also remain independent of the dissipation rate of the battery $\Gamma_{B}$. Furthermore, when the dissipation rate $\Gamma_{B}$ of the quantum battery increases to a certain value, the ergotropy of the battery is zero regardless of the value of $\Gamma_{C}$ and $\delta$. This indicates that no energy can be extracted from the battery, and the feedback control mechanism becomes ineffective.

Under the optimal charging parameters ($\Gamma_{C}=2g$ and $\delta=f/\Gamma_{C}=1$), the maximum extractable energy (the ergotropy) of the quantum battery in the bosonic thermal reservoir is
\begin{equation}
\mathcal{E}(\infty)=\omega_{0}\left[\frac{8g^2(1+n_{b})+(1+2n_{b})\lvert\beta_{b}\lvert}{2(1+2n_{b})(2g+\Gamma_{B}+2n_{b}\Gamma_{B})^2}-\frac{1}{2(1+2n_{b})}\right],
\end{equation}
where $\beta_{b}=-4g^2+4g\Gamma_{B}+(1+2n_{b})\Gamma_{B}^2$. Here, $\beta_{b}$ can be positive or negative; therefore, we will discuss the cases separately. When $\beta_{b}\geq 0$, the ergotropy of the battery becomes $\mathcal{E}(\infty)=0$, indicating that no energy can be extracted from the battery. According to the condition of $\beta_{b}=0$, we can also determine that the critical value of the battery dissipation rate $\Gamma_{B}$ at different temperatures is $\Gamma_{B}/g=2\left(-1+\sqrt{2}\sqrt{1+n_{b}}\right)/(1+2n_{b})$. When the dissipation rate of the battery $\Gamma_{B}$ exceeds this critical value, no energy can be extracted from the battery. Furthermore, the result shows that this critical value decreases monotonically with increasing temperature. Under high-temperature conditions, the quantum battery must maintain a sufficiently low dissipation rate to ensure the viability of the charging protocol; otherwise, the protocol fails. These results highlight the fundamental limitations that temperature imposes on the quantum battery charging protocol. When $\beta_{b}<0$, the ergotropy of the quantum battery becomes
\begin{equation}
\mathcal{E}(\infty)=\omega_{0}\frac{4g^2-4g\Gamma_{B}-(1+2n_{b})\Gamma_{B}^2}{(2g+\Gamma_{B}+2n_{b}\Gamma_{B})^2}.
\end{equation}
The result indicates that as the temperature increases, the extractable energy of the battery gradually decreases. In general, when the battery is in a bosonic thermal reservoir, the temperature not only narrows the permissible range of the dissipation rate of the battery but also reduces its extractable energy.
\begin{figure}
\centering\scalebox{0.36}{\includegraphics{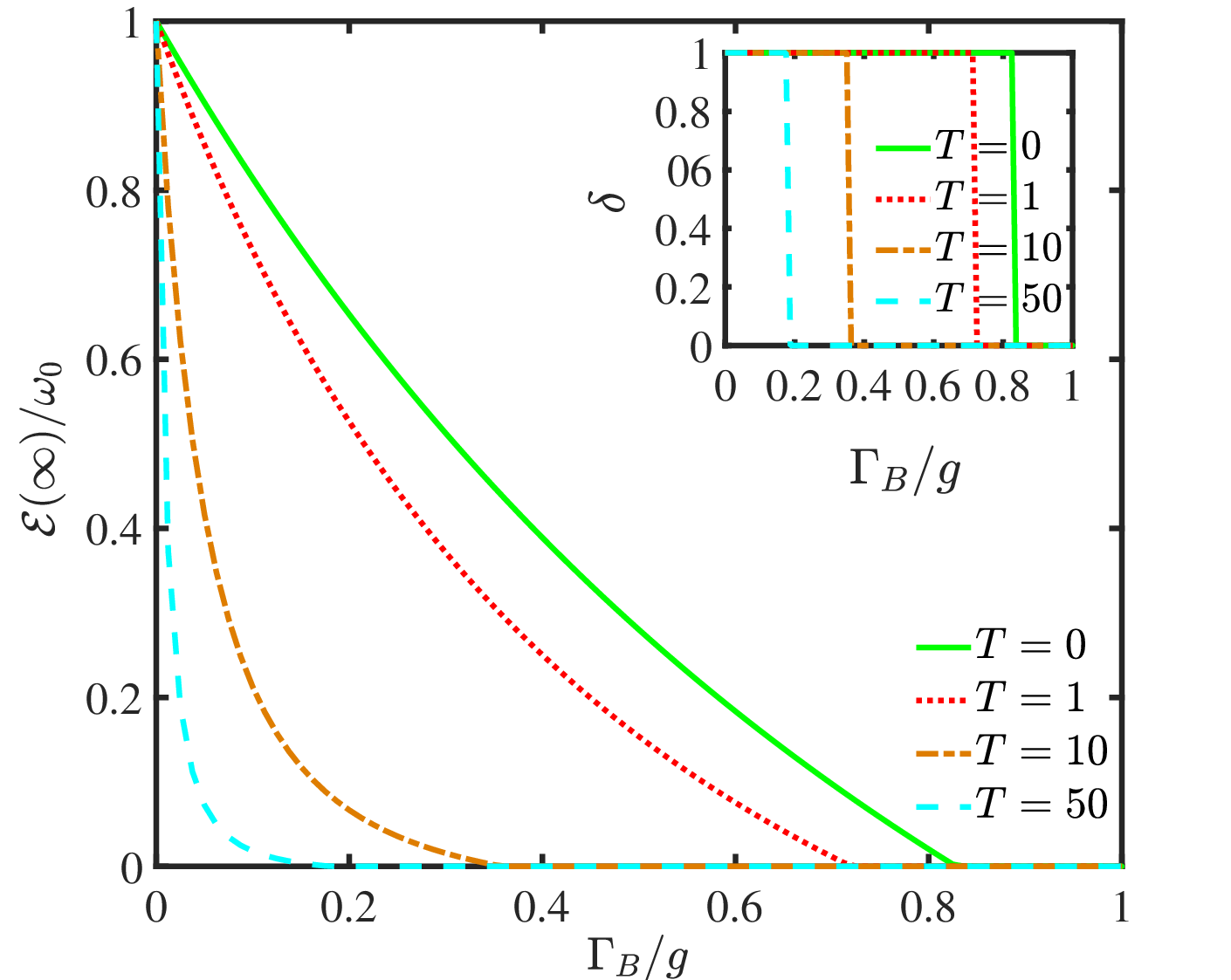}}
\caption{\label{Q7} Ergotropy of the quantum battery in the steady state plotted as a function of the battery dissipation rate $\Gamma_{B}$, where the dissipation rate of the charger is fixed at its optimal value, i.e., $\Gamma_{C}=2g$. The corresponding optimal feedback parameter $\delta$ is shown in the inset. Different curves correspond to different thermal reservoir temperatures. The green solid line, red dotted line, yellow dash-dotted line, and blue dashed line correspond to bosonic thermal reservoir temperatures $T$ of 0, 1, 10, and 50, respectively. The remaining parameters are the same as in Fig.~\ref{Q2}.
 }
\end{figure}

Subsequently, to more intuitively illustrate the impact of temperature and battery dissipation rate on the ergotropy of the quantum battery, we further analyze how the ergotropy of the quantum battery varies with the dissipation rate $\Gamma_{B}$ of the battery at different temperatures, as shown in Fig.~\ref{Q7}, where the dissipation rate of the charger is set to the optimal value $\Gamma_{C}=2g$. The corresponding optimal feedback parameters $\delta$ are also shown in the inset. The results show that when the temperature of the thermal reservoir is constant, the ergotropy gradually decreases with increasing dissipation rate $\Gamma_{B}$. At the same time, the dissipation rate $\Gamma_{B}$ of the battery has a critical value. If $\Gamma_{B}$ exceeds this threshold, the battery extraction energy becomes zero. In addition, this critical value decreases as the temperature increases. The inset shows that if $\Gamma_{B}$ does not exceed this critical value, the corresponding optimal feedback parameter remains $\delta=1$. Furthermore, it can be observed that increasing the temperature is not conducive to extracting energy from the battery.
\begin{figure}
\centering\scalebox{0.36}{\includegraphics{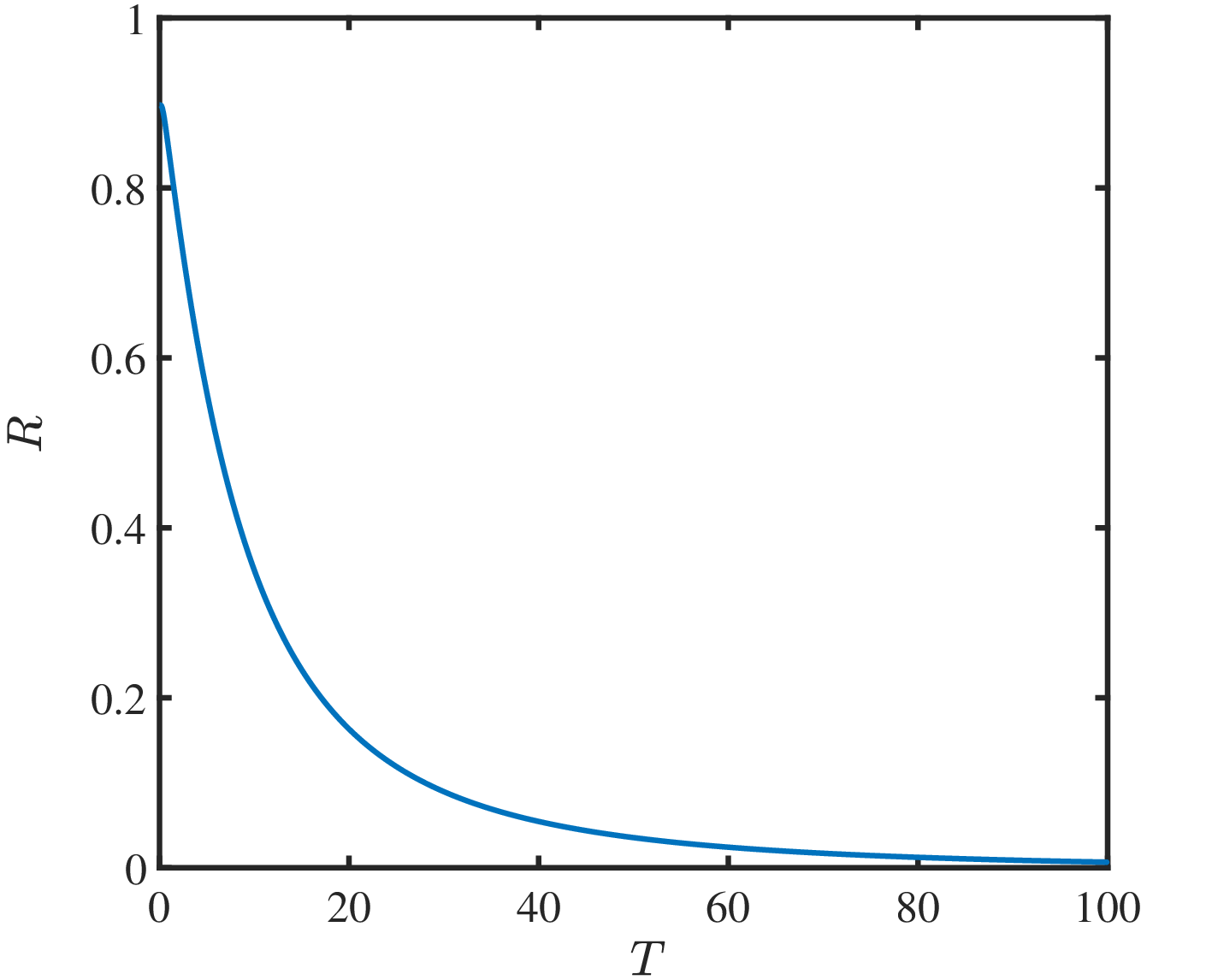}}
\caption{\label{Q8} Under the optimal charging conditions in a bosonic thermal reservoir, the charging efficiency $R$ of the quantum battery is plotted as a function of temperature $T$, with $\Gamma_{B}=0.1g$.
 }
\end{figure}

To evaluate the influence of the environment of the bosonic thermal reservoir on the battery charging efficiency $R$, we explore the impact of the temperature of the bosonic thermal reservoir on the quantum battery charging efficiency under optimal charging conditions ($\Gamma_{C}=2g$ and $\delta=1$), as shown in Fig.~\ref{Q8}, with $\Gamma_{B}=0.1g$. The definition of battery charging efficiency is given in Eq.~(\ref{R}). The results show that the charging efficiency $R$ of the quantum battery decreases with increasing temperature $T$, which means that the increase in the temperature of the reservoir inhibits the efficiency of energy extraction from the battery.

\subsection{Charging process of quantum battery in a fermionic thermal reservoir}
Next we consider the scenario in which the quantum battery is embedded in a finite-temperature fermionic thermal reservoir. By solving Eq.~(\ref{master}), we obtain that the stored energy of the quantum battery is given by
\begin{equation}
E_{B}(\infty)=\omega_{0}\frac{4g^{2}Q_{f}(\Gamma_{B}\eta+\Gamma_{C}S)+n_{f}W_{f}S}{4g^2(\Gamma_{B}\eta+\Gamma_{C}S)^{2}+W_{f}S},
\end{equation}
where we set $\delta=f/\Gamma_{C}$, $S=2\delta(\delta-\eta)+\eta$, $Q_{f}=\Gamma_{C}\delta^{2}+n_{f}\Gamma_{B}\eta$, and $W_{f}=\Gamma_{C}\Gamma_{B}(\Gamma_{C}+\Gamma_{B})[\Gamma_{B}\eta+\Gamma_{C}(2S-\eta)]$.
\begin{figure}
\centering\scalebox{0.27}{\includegraphics{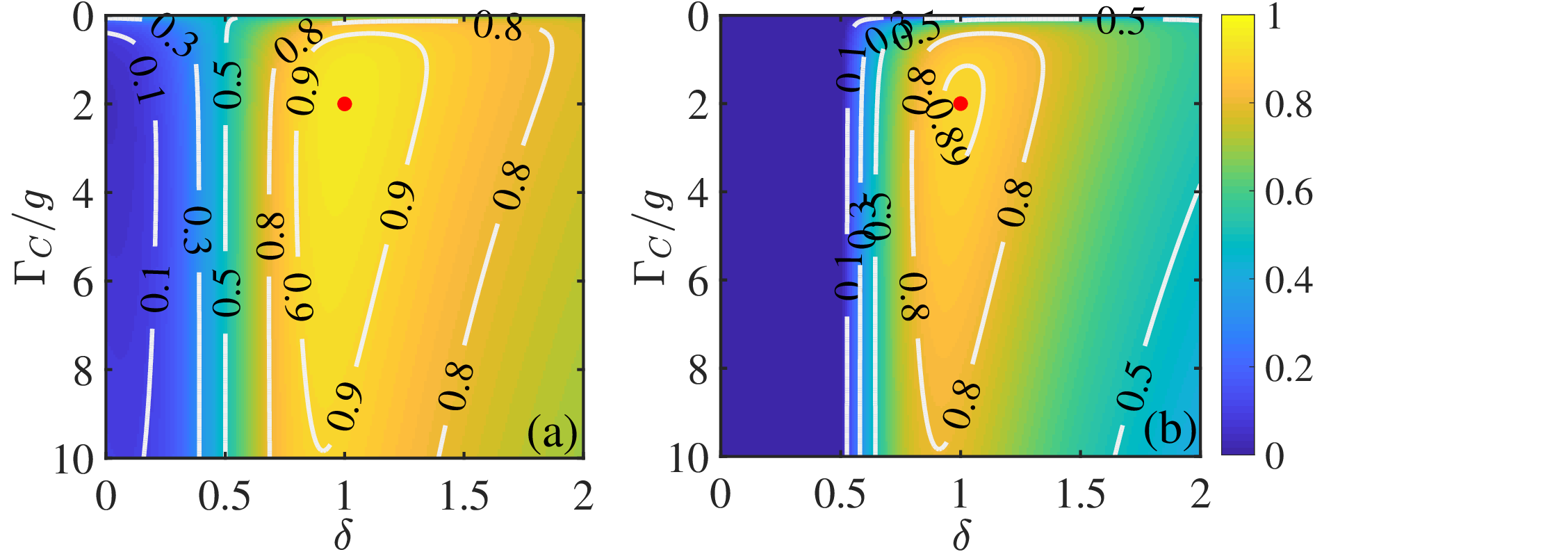}}
\caption{\label{Q9} The quantum battery is placed in a fermionic reservoir. Plotted are (a) the steady-state stored energy and (b) the ergotropy of the quantum battery, as a function of the feedback parameter $\delta$ and the dissipation rate $\Gamma_{C}$ of the charger. The temperature of the fermionic reservoir is $T=10$ and the other parameters are the same as in Fig.~\ref{Q2}.
 }
\end{figure}

\begin{figure}
\centering\scalebox{0.27}{\includegraphics{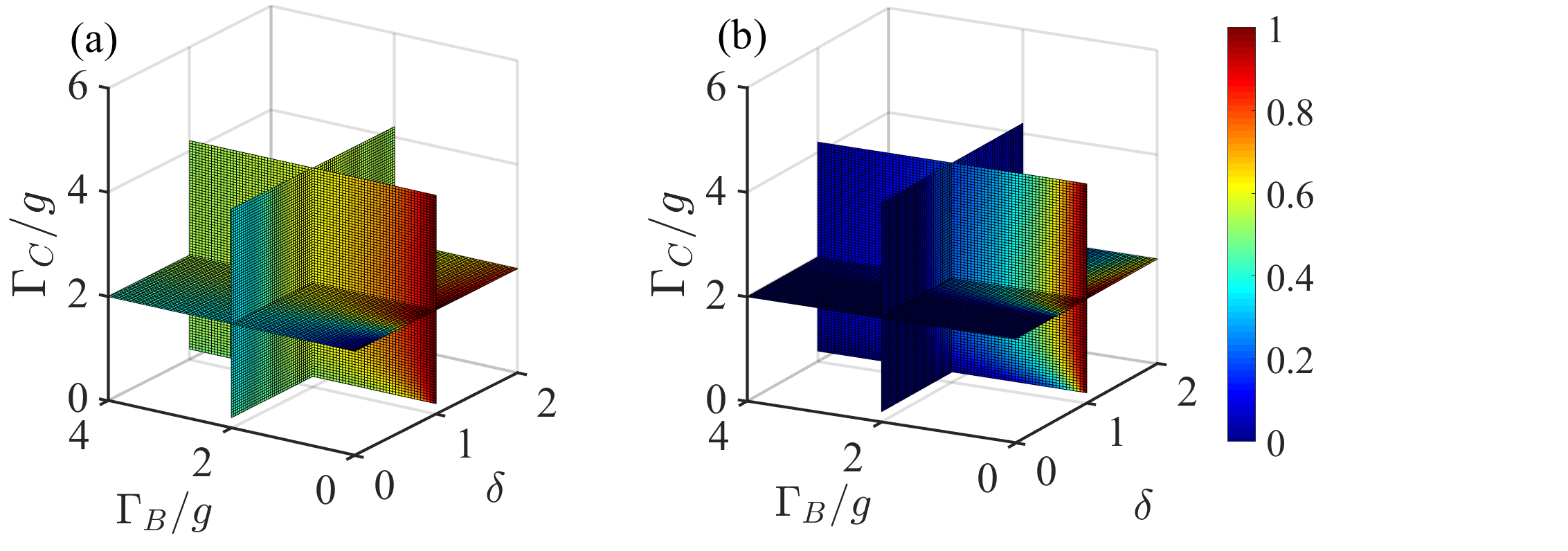}}
\caption{\label{Q10} (a) Steady-state stored energy and (b) ergotropy of the quantum battery, as a function of $\delta$, $\Gamma_{B}$, and $\Gamma_{C}$. The fermionic reservoir temperature is set to $T=10$. All other parameters are the same as in Fig.~\ref{Q2}.
 }
\end{figure}

For a zero-temperature thermal reservoir, the average excitation number of the thermal reservoir satisfies $n_{b(f)}=0$. As Eq.~(\ref{master}) shows, the master equation governing the dynamics of the control system has the same form for both bosonic and fermionic thermal reservoirs. Therefore, we will not specifically consider the scenario in which the temperature of the fermionic thermal reservoir is $T=0$. Figures~\ref{Q9}(a) and \ref{Q9}(b) show the steady-state dependence of the battery's stored energy and ergotropy on the feedback parameter $\delta$ and the charger dissipation rate $\Gamma_{C}$, respectively. Here the temperature of the thermal reservoir is set at $T=10$ and the dissipation rate of the battery is set at $\Gamma_{B}=0.1g$. The results reveal the existence of optimal charging parameters that simultaneously maximize both the stored energy and the ergotropy of the quantum battery, which coincide with those obtained for the bosonic reservoir case. Furthermore, Fig.~\ref{Q10} again confirms that these parameters remain independent of the battery dissipation rate $\Gamma_{B}$.

\begin{figure}
\centering\scalebox{0.36}{\includegraphics{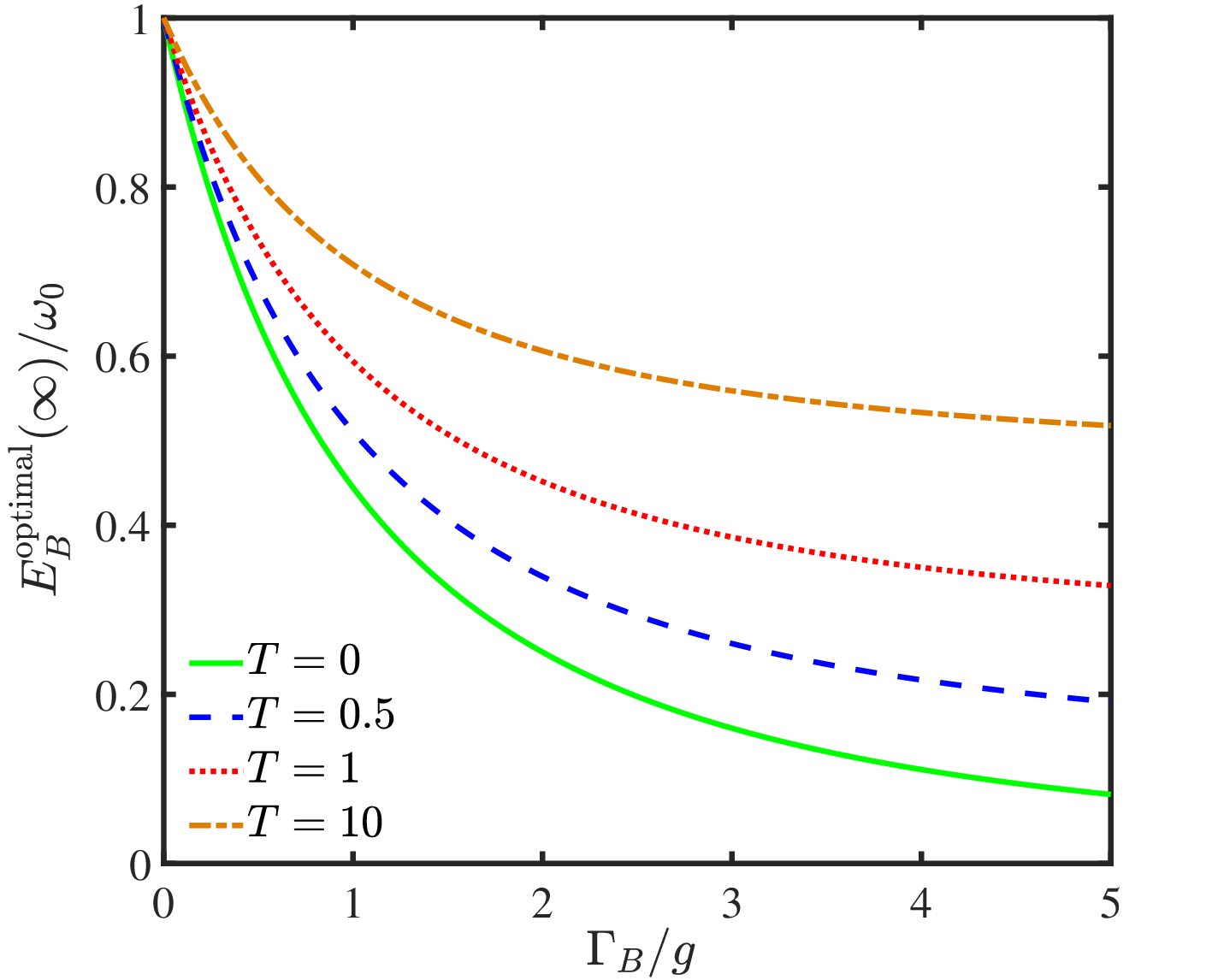}}
\caption{\label{Q11} Under optimal charging conditions $\Gamma_{C}=2g$ and $\delta=1$, the steady-state stored energy of the quantum battery is plotted as a function of the battery dissipation rate. Different curves correspond to different reservoir temperatures. The green solid line, blue dashed line, red dotted line, and yellow dash-dotted line correspond to the fermionic thermal reservoir temperatures $T$ of 0, 0.5, 1, and 10, respectively. The average thermal excitation number follows $n_{f}=1/(e^{1/T}+1)$, where $T=0$ corresponds to $n_{f}=0$. The other parameters match those in Fig.~\ref{Q2}.
 }
\end{figure}

Under the perfect measurement ($\eta=1$) and the optimal charging conditions ($\Gamma_{C}=2g$ and $\delta=1$), the stored energy of the battery becomes
\begin{equation}\label{e16}
E_{B}^{\rm optimal}(\infty)=\omega_{0}\frac{4g^{2}+n_{f}(\Gamma_{B}^{2}+4g\Gamma_{B})}{(2g+\Gamma_{B})^{2}}.
\end{equation}
As shown in Eq.~(\ref{e16}), for a fixed dissipation rate $\Gamma_{B}$, the stored energy of the quantum battery increases monotonically with temperature. The corresponding numerical results appear in Fig.~\ref{Q11}, with distinct curves corresponding to different reservoir temperatures. In contrast to the bosonic reservoir case, elevated temperatures consistently enhance the battery energy storage capacity, regardless of the magnitude of the dissipation rate $\Gamma_{B}$ of the battery.
\begin{figure}
\centering\scalebox{0.36}{\includegraphics{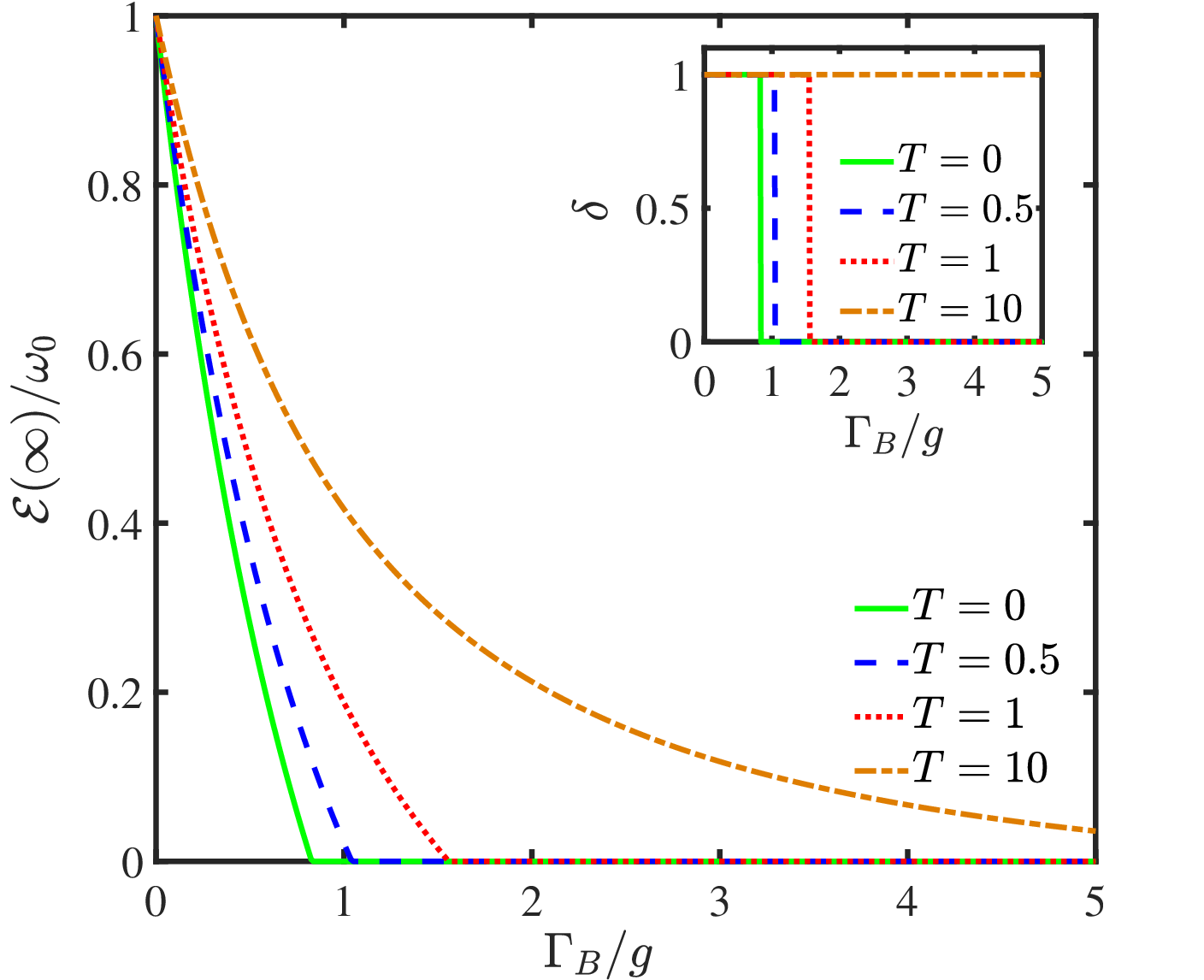}}
\caption{\label{Q12} Steady-state ergotropy of the quantum battery plotted as a function of $\Gamma_{B}$, with different curves corresponding to different fermionic thermal reservoir temperatures. The dissipation rate of the charger is set to $\Gamma_{C}=2g$ and the corresponding optimal feedback parameters $\delta$ are shown in the inset.
 }
\end{figure}

Under optimal charging parameters ($\Gamma_{C}=2g$ and $\delta=1$), the maximum extractable energy (the ergotropy) of the quantum battery can be expressed as
\begin{eqnarray}
\mathcal{E}(\infty)&=&\omega_{0}\frac{\beta_{f}+\lvert\beta_{f}\rvert}{2(2g+\Gamma_{B})^2},
\end{eqnarray}
where $\beta_{f}=4g^2 + 4g(-1+2n_{f})\Gamma_{B} + (-1+2n_{f})\Gamma_{B}^2$. We analyze the problem in two scenarios. First, when $\beta_{f}\leq0$, we obtain $\mathcal{E}(\infty)=0$, which means that energy cannot be extracted from the quantum battery. Setting $\beta_{f}=0$, we derive the critical battery dissipation rate in the positive temperature regime as $\Gamma_{B}/g=2/(1-2n_{f}+\sqrt{4n_{f}^{2}-6n_{f}+2})$, which is a temperature-dependent function. Once $\Gamma_{B}$ exceeds the critical value, the battery can no longer supply energy to other devices. On the basis of the critical condition analysis, we observe that as the temperature increases, the permissible range of the battery dissipation rate broadens. When $\beta_{f}>0$, then
\begin{eqnarray}
\mathcal{E}(\infty)&=&\omega_{0}\frac{2\beta_{f}}{2(2g+\Gamma_{B})^2}\nonumber\\&=&\omega_{0}\left[\frac{8g\Gamma_{B}+2\Gamma_{B}^{2}}{(2g+\Gamma_{B})^2}n_{f}+\frac{4g^{2}-4g\Gamma_{B}-\Gamma_{B}^{2}}{(2g+\Gamma_{B})^2}\right].
\end{eqnarray}

\begin{figure}
\centering\scalebox{0.36}{\includegraphics{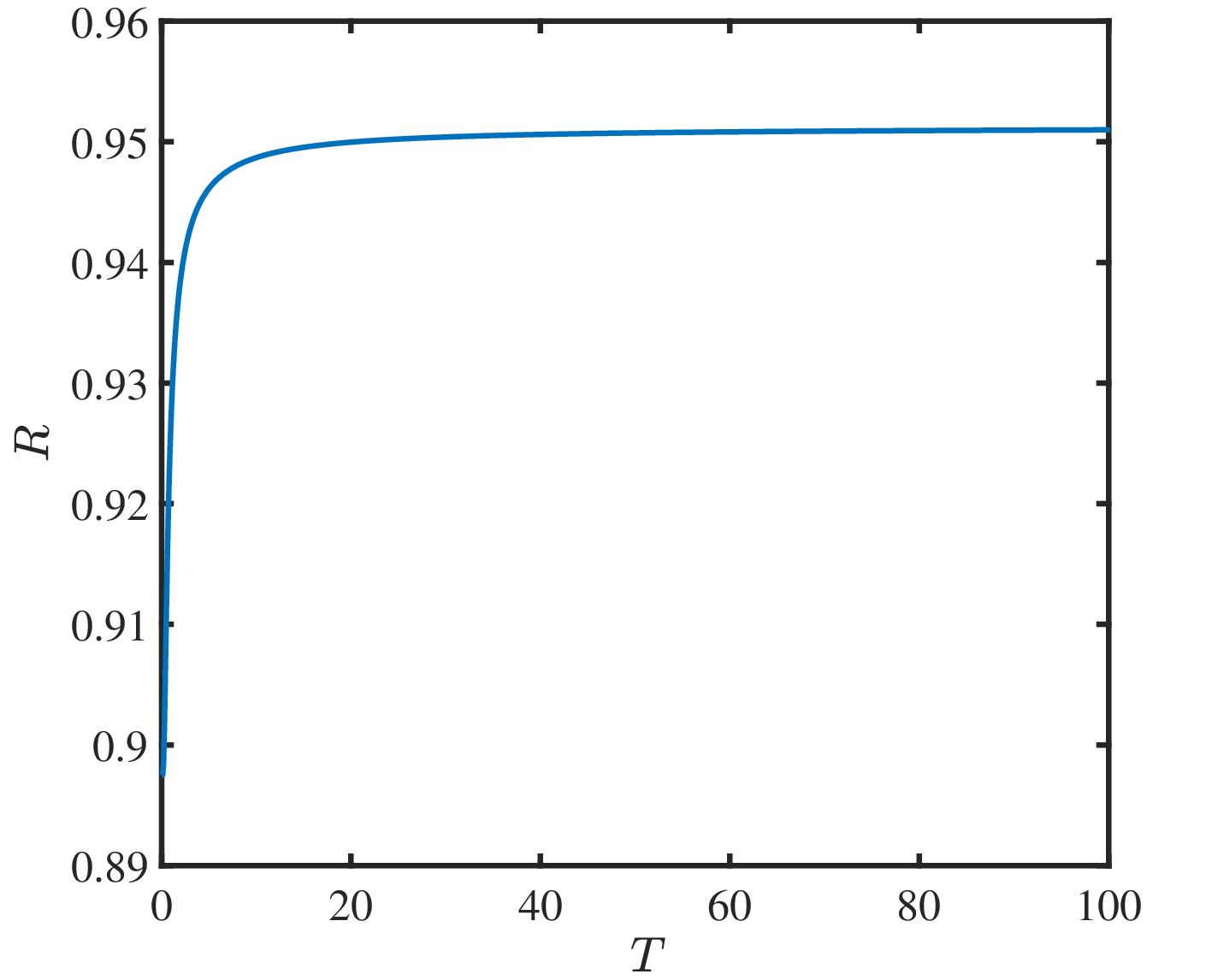}}
\caption{\label{Q13} Under the optimal charging conditions of the fermionic thermal reservoir, the charging efficiency $R$ of the quantum battery is plotted as a function of the temperature $T$, with $\Gamma_{B}=0.1g$.
 }
\end{figure}
This equation indicates that, for fixed $g$ and $\Gamma_{B}$, the extractable energy from the quantum battery increases with temperature. In summary, when the quantum battery is coupled to a fermionic thermal reservoir, higher temperatures not only broaden the permissible range of the battery's dissipation rate but also enhance the battery's energy extraction capability.
\begin{figure*}
\centering\scalebox{0.4}{\includegraphics{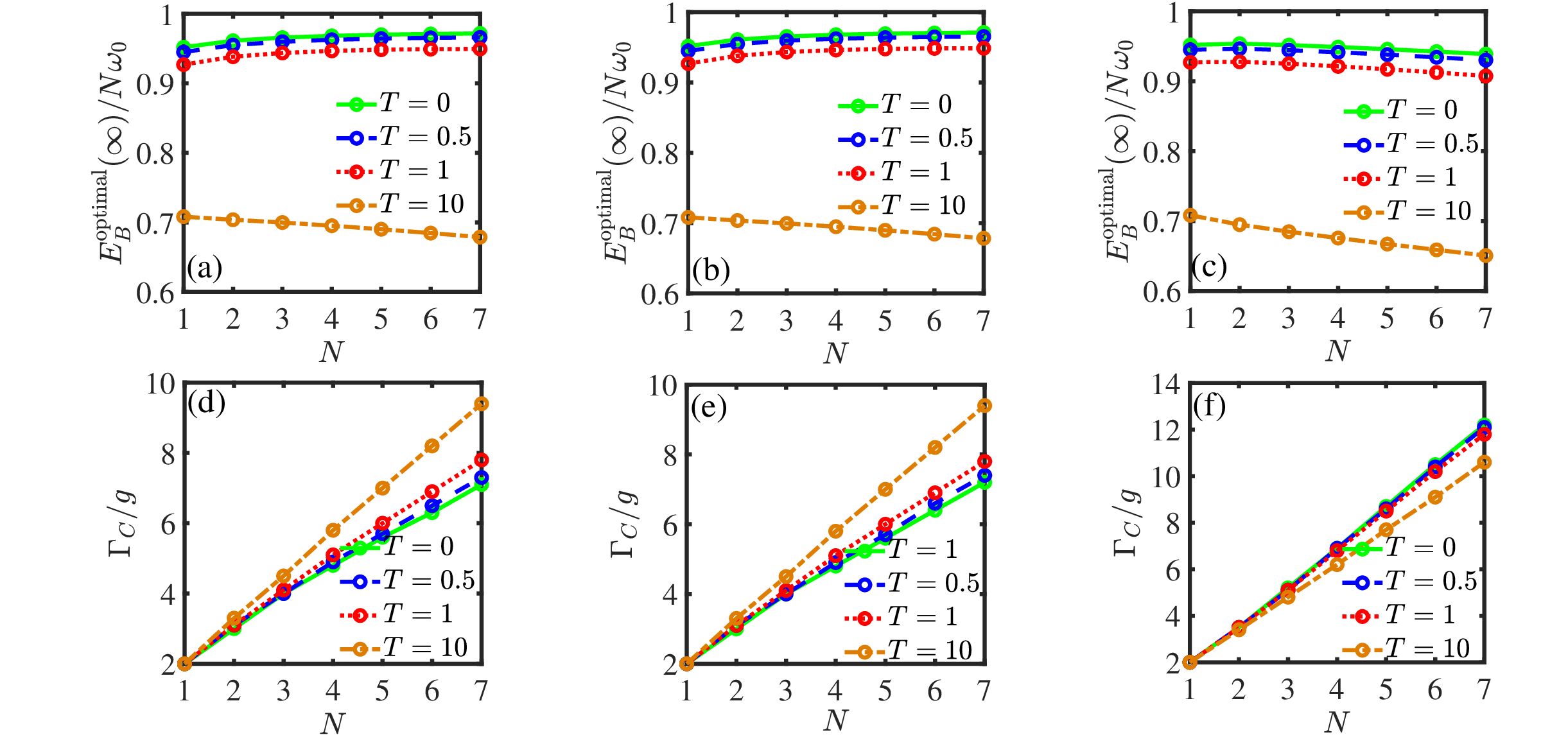}}
\caption{\label{BM1} The multiparticle quantum battery is placed in a bosonic thermal reservoir, and its optimal energy storage is analyzed as a function of the number of constituent quantum battery particles at different reservoir temperatures. The coupling strengths between the internal particles of the quantum battery are (a) $J=0$, (b) $J=0.1g$, and (c) $J=g$. The corresponding optimal charger dissipation parameters $\Gamma_{C}$ are shown in (d)-(f). The remaining parameters are set as $g=0.01\omega_{0}$, $\delta=f/\Gamma_{C}=1$, and $\Gamma_{B}=0.05g$.
 }
\end{figure*}

Figure~\ref{Q12} illustrates the dependence of the quantum battery ergotropy on its dissipation rate, where the dissipation rate of the charger remains at its optimal value ($\Gamma_{C}=2g$). The corresponding optimal feedback parameters are presented in the inset. The results demonstrate that increasing temperature promotes the extraction of battery energy. Similarly to the bosonic thermal reservoir case, there exists a critical value for the dissipation rate $\Gamma_{B}$ of the battery in terms of energy extraction. In particular, when the battery is coupled to a fermionic thermal reservoir, an increase in temperature extends the critical point of the dissipation rate of the battery $\Gamma_{B}$ and improves its usability. Subsequently, we explore the influence of the temperature of the fermionic thermal reservoir on the charging efficiency $R$ of the battery under optimal charging conditions through Fig.~\ref{Q13}, where the dissipation rate of the battery is $\Gamma_{B}=0.1g$. The results show that the charging efficiency $R$ increases with temperature, suggesting that the temperature can serve as a resource to improve battery performance. This is different from the scheme proposed in Ref.~\cite{Song_2024}, which exhibits the temperature will be unfavorable for the extraction of battery energy, when the environment is within the positive temperature range, while in our scheme, temperature is still a favorable factor for battery charging. We conduct a simple comparative analysis of these two schemes in Appendix~\ref{B}.

\section{MULTIPARTICLE QUANTUM BATTERY MODEL}\label{IV}
On the basis of the quantum feedback mechanism, we extend the previously proposed single-particle battery model to a multiparticle configuration. In this extended model, the charger consists of a single two-level atom embedded in a zero-temperature reservoir. During the charging process, quantum feedback control is applied to the charger. The quantum battery is composed of multiple two-level atoms and is embedded in a common thermal reservoir. The coupling strength between the charger and each atom in the battery is $g$. The free Hamiltonian form of the multiparticle quantum battery is given as
\begin{equation}
H_{B}=\frac{\omega_{0}}{2}\sum_{i=1}^{N}\sigma_{iB}^{z}=\omega_{0}S_{B}^{z},
\end{equation}
where $S_{B}^{x,y,z}=\sum_{i}\sigma_{iB}^{x,y,z}/2$ defines the collective atom operators of the quantum battery and $N$ denotes the number of atoms comprising the battery.

In the Markovian limit (i.e.,$\tau\rightarrow0$), the evolution of the system is described by the master equation
\begin{eqnarray}\label{master1}
\mathop{\dot{\rho}}&=&-i[H_{\rm SI}',\rho]+if[\sigma_{C}^{y},\sigma_{C}^{-}\rho+\rho\sigma_{C}^{+}]+\frac{f^{2}}{\eta\Gamma_{C}}\mathcal{D}[\sigma_{C}^{y}]\rho+\mathcal{L}_{C}\rho\nonumber\\&&+\mathcal{L}_{B}'\rho.
\end{eqnarray}
At this point, the interaction term $H_{SI}'$ between the charger and the quantum battery, and the Lindblad term $\mathcal{L}_{B}'\rho$ for the quantum battery, are given in the forms
\begin{equation}\label{HSI'}
H_{SI}'=\sum_{i=1}^{N}g_{i}(\sigma_{C}^{+}\sigma_{iB}^{-}+\sigma_{C}^{-}\sigma_{iB}^{+})+\sum_{i<j}^{N}J_{i}(\sigma_{iB}^{+}\sigma_{jB}^{-}+\sigma_{iB}^{-}\sigma_{jB}^{+}).
\end{equation}
\begin{eqnarray}\label{LB}
\mathcal{L}_{B}'\rho &=&\gamma^{\downarrow}_{B}[L_{B}^{-}\rho L_{B}^{+}-1/2(L_{B}^{+}L_{B}^{-}\rho+\rho L_{B}^{+}L_{B}^{-})]\nonumber\\&&+\gamma^{\uparrow}_{B}[L_{B}^{+}\rho L_{B}^{-}-1/2(L_{B}^{-}L_{B}^{+}\rho+\rho L_{B}^{-}L_{B}^{+})].
\end{eqnarray}
The first term of Eq.~(\ref{HSI'}) describes the direct coupling between the charger and the constituent atoms of the quantum battery, while the second term accounts for the interactions between the atoms within the battery itself. For simplicity, we assume $g_{i}=g$ and $J_{i}=J$. In Eq.~(\ref{LB}), $L_{B}^{-}=\sum_{i=1}^{N}\sigma_{iB}^{-}$ defines the collective lowering operator of the quantum battery. To facilitate a fair comparison with the single-particle quantum battery, the steady-state energy density of the multiparticle quantum battery is defined as
\begin{equation}
\mathcal{W}=\frac{E_{B}(\infty)}{N}=\omega_{0}\left(\frac{\langle S_{B}^{z}(\infty)\rangle}{N}+\frac{1}{2}\right ).
\end{equation}
Thus, the steady-state stored energy of the battery is
\begin{equation}
E_{B}(\infty)=N\mathcal{W}.
\end{equation}
\begin{figure*}
\centering\scalebox{0.4}{\includegraphics{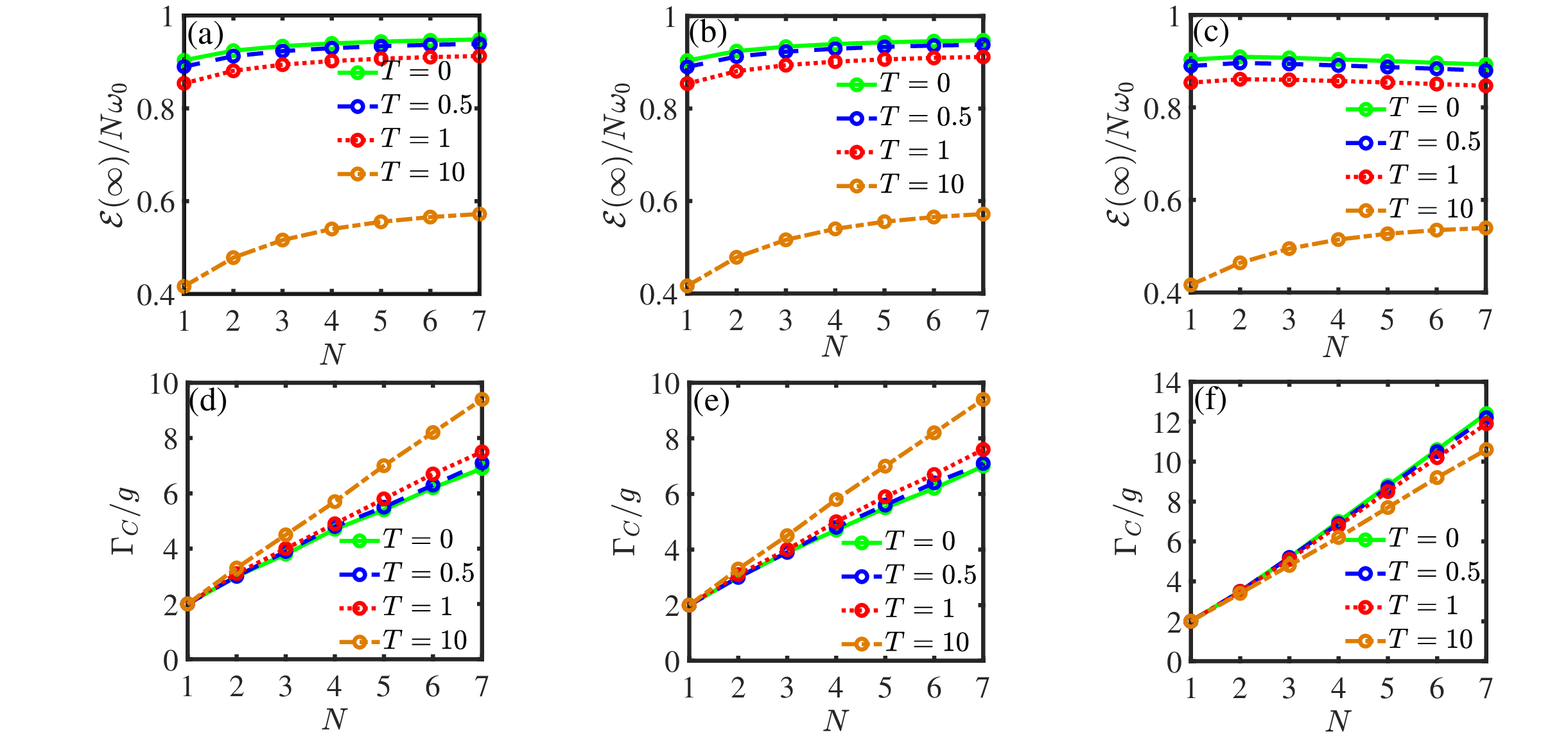}}
\caption{\label{BM2} In the bosonic thermal reservoir, the variation of the average ergotropy of the quantum battery in the steady state is plotted as a function of the particle number $N$, with the corresponding atom-atom coupling strengths in the multiparticle quantum battery set to (a) $J=0$, (b) $J=0.1g$, and (c) $J=g$. (d)-(f) Optimal dissipation rate $\Gamma_{C}$ of the charger corresponding to (a)-(c). Different curves represent different temperatures, the other parameters are identical to those in Fig.~\ref{BM1}.
 }
\end{figure*}
\begin{figure*}
\centering\scalebox{0.4}{\includegraphics{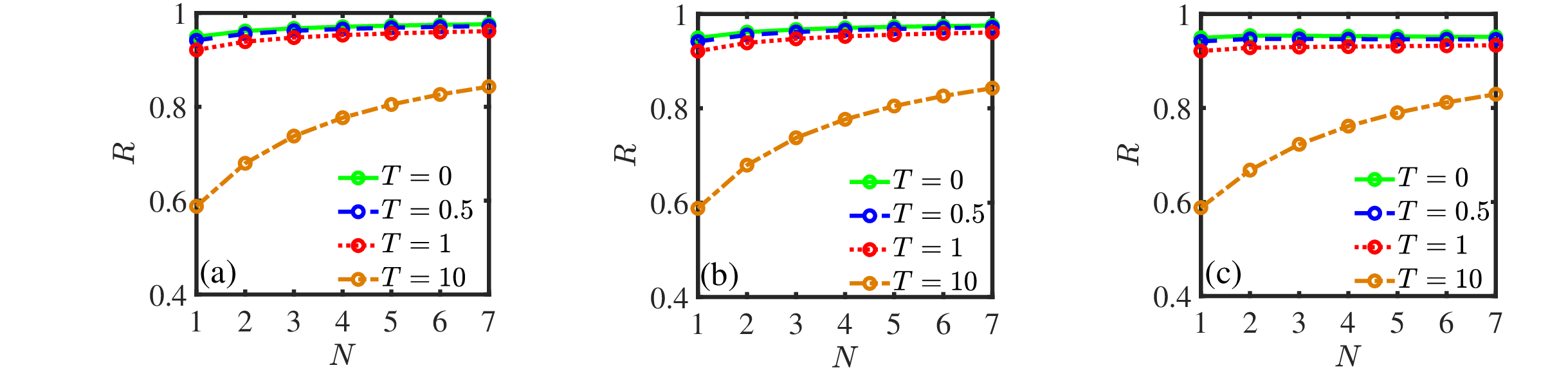}}
\caption{\label{BM3} The multiparticle quantum battery is placed in a bosonic thermal reservoir. The charging efficiency $R$ of the battery is analyzed as a function of the number of particles $N$. The interparticle interaction strengths of the multiparticle quantum battery are set to (a) $J=0$, (b)  $J=0.1g$, and (c) $J=g$. Different curves represent different temperatures; all other parameters are identical to those in Fig.~\ref{BM1}.
 }
\end{figure*}

In the following, we investigate the performance of the multiparticle quantum battery under different thermal reservoirs. The feedback parameter is set to $\delta=f/\Gamma_{C}=1$ to facilitate a more meaningful comparison with the single-particle quantum battery in the subsequent analysis.
\subsection{Charging process of the multiparticle quantum battery in a bosonic thermal reservoir}
First, we investigate the charging process of the multiparticle quantum battery in a bosonic thermal reservoir, in this case, $\gamma^{\downarrow}_{B}=\Gamma_{B}(1+n_{b})$ and $\gamma^{\uparrow}_{B}=\Gamma_{B}n_{b}$ in Eq.~(\ref{LB}). To explore the effects of the number of quantum battery particles $N$ and the interparticle interactions $J$ on battery performance, we examine the optimal energy density of the battery as a function of $N$ under three coupling regimes: no coupling ($J=0$), weak coupling ($J=0.1g$), and strong coupling ($J=g$), as shown in Figs.~\ref{BM1}(a)-\ref{BM1}(c). The quantum battery operates in the desired weak dissipation regime with $\Gamma_{B}=0.05g$, and the different curves correspond to different thermal reservoir temperatures. Additionally, the corresponding optimal charger dissipation rates for Figs.~\ref{BM1}(a)-\ref{BM1}(c) are presented in Figs.~\ref{BM1}(d)-\ref{BM1}(f). The results indicate that, similar to the single-particle quantum battery case, temperature remains a detrimental factor for charging. Furthermore, it is observed that, under low-temperature conditions, the energy density of the battery increases with the number of particles when the interparticle couplings are absent or weak, as shown in Figs.~\ref{BM1}(a) and \ref{BM1}(b). However, at higher temperatures, the energy density decreases as particle number increases, thus impairing efficient energy transfer to the battery. Furthermore, in the strong interparticle coupling regime, both elevated temperature and an increased number of particles have detrimental effects on the energy storage capacity of the battery, as illustrated in Fig.~\ref{BM1}(c).
\begin{figure}
\centering\scalebox{0.33}{\includegraphics{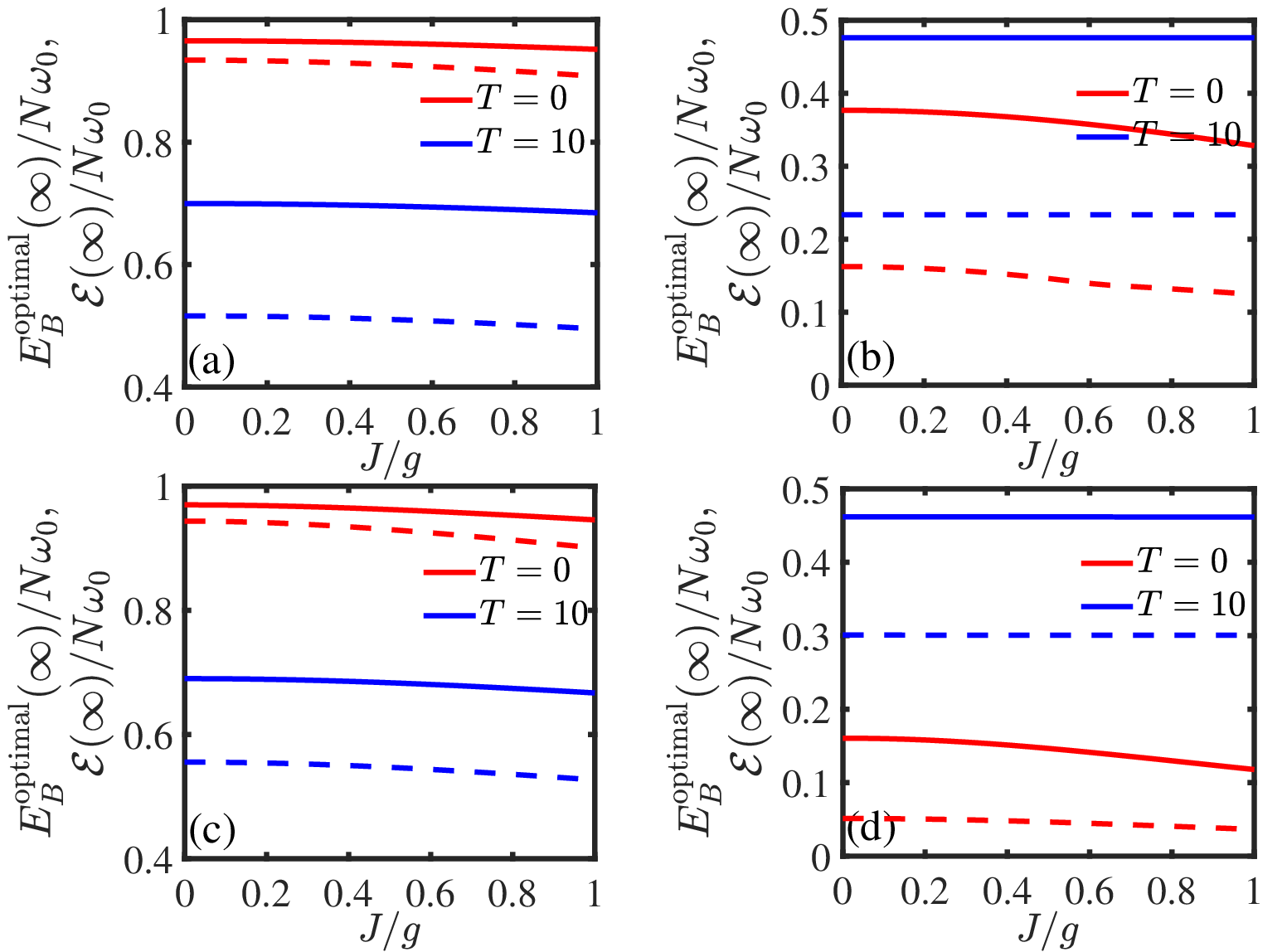}}
\caption{\label{BM4} The multiparticle quantum battery is placed in a bosonic thermal reservoir. The solid line represents the energy density of the multiparticle quantum battery, while the dashed line represents the maximum extractable energy per particle (average ergotropy). Curves of different colors correspond to different temperatures, with the red curve representing $T=0$ ($n_{b}=0$) and the blue curve representing $T=10$ [$n_{b}=1/(e^{1/10}-1)\approx9.51$]. Plotted are three-particle quantum batteries with dissipation rates (a) $\Gamma_{B}=0.05g$ and (b) $\Gamma_{B}=0.5g$ and five-particle quantum batteries with dissipation rates (c) $\Gamma_{B}=0.05g$ and (d) $\Gamma_{B}=0.5g$. All other parameters are identical to those in Fig.~\ref{BM1}.
 }
\end{figure}
\begin{figure*}
\centering\scalebox{0.4}{\includegraphics{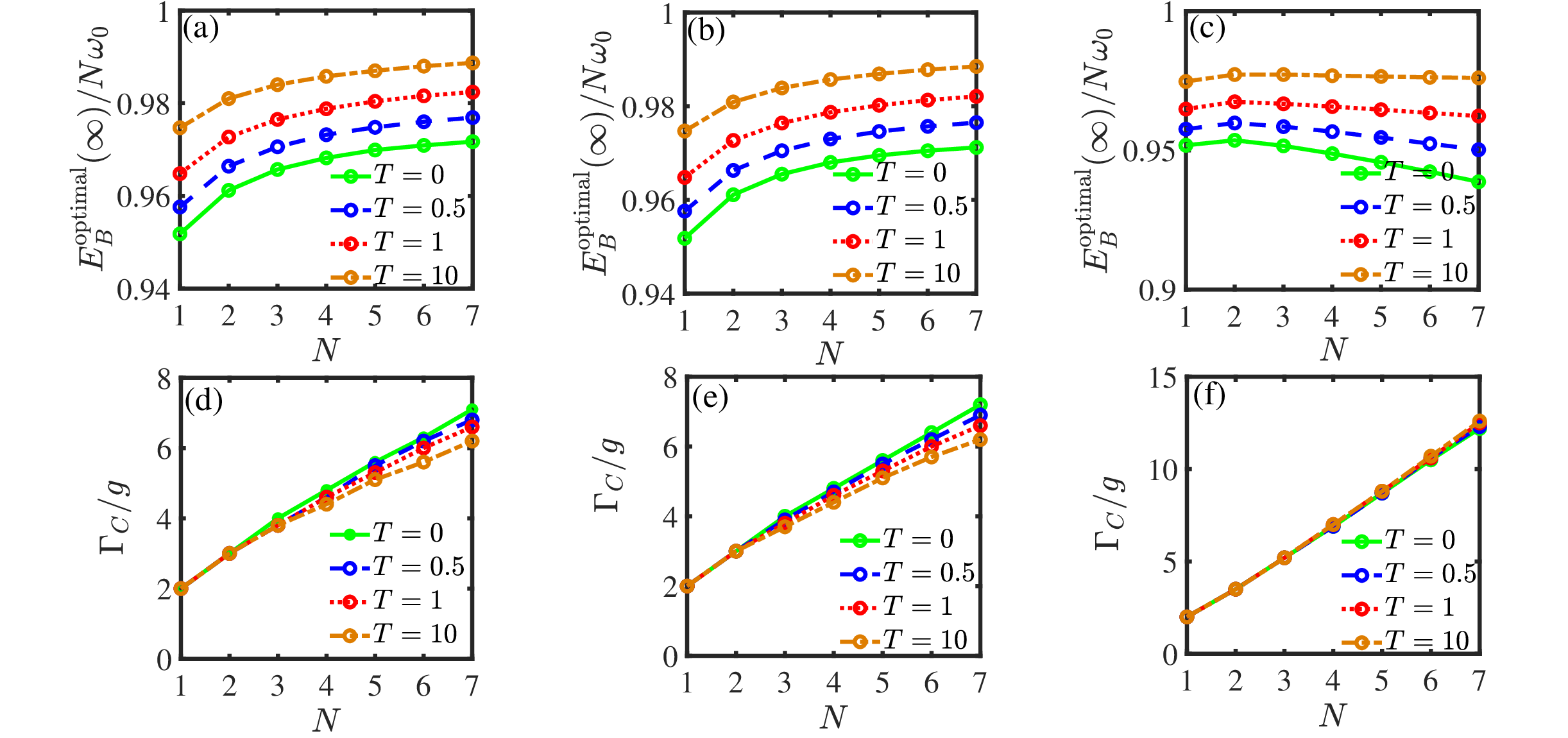}}
\caption{\label{FM1} The multiparticle quantum battery is placed in a fermionic thermal reservoir. The depict the variation in the steady-state energy density of the multiparticle quantum battery is plotted as a function of particle number $N$ for (a) $J=0$, (b) $J=0.1g$, and (c) $J=g$. (d)-(f) Optimal charger dissipation parameters required for (a)-(c). Different curves correspond to different temperatures. The other parameters are the same as those in Fig.~\ref{BM1}.
 }
\end{figure*}

Subsequently, Fig.~\ref{BM2} presents the dependence of the maximum extractable energy per particle (i.e., the average ergotropy) of the multiparticle quantum battery on the number of particles $N$. As shown in Figs.~\ref{BM2}(a)-\ref{BM2}(b), when there is no interparticle interaction ($J=0$) or the interaction is weak ($J=0.1g$), the average ergotropy of the battery increases as the number of particles $N$ increases. Meanwhile, when the number of particles in a multiparticle quantum battery is fixed, the average ergotropy decreases as the temperature increases, which means that temperature inhibits the extraction of energy from the battery. In addition, when the interaction between particles in a multiparticle quantum battery is strong ($J=g$), the average ergotropy of the battery exhibits a slight decrease with increasing particle number $N$ at low temperatures. In contrast, at high temperatures, this behavior is suppressed, and the average ergotropy increases as the number of particles $N$ increases, but remains lower than that observed at low temperatures. As seen in Figs.~\ref{BM1}(d)-\ref{BM1}(f) and \ref{BM2}(d)-\ref{BM2}(f), in terms of both energy storage and energy extraction, the optimal dissipation rate of the charger required increases with the number of particles. Figure~\ref{BM3} investigates the effects of temperature and particle number $N$ on the charging efficiency $R$ of the battery. Figures~\ref{BM3}(a)–\ref{BM3}(c) correspond to $J=0$, $0.1g$, and $g$, respectively, and the different curves represent different temperatures of the thermal reservoir. It is observed that at high temperatures, the charging efficiency $R$ is more significantly affected by the number of battery particles, whereas this influence is greatly reduced at low temperatures. Overall, temperature remains a detrimental factor for both energy storage and extraction in the multiparticle quantum battery. However, there exist two specific scenarios in which an increase in the number of quantum battery particles can be beneficial for energy extraction: One occurs with weak interparticle interactions, and the other at high temperatures with strong interparticle interactions. The multiparticle quantum battery operates in a strong dissipation regime, as discussed in the Appendix~\ref {C}.

In the following, we further investigate the effect of interparticle interaction strength $J$ on the energy density and the average ergotropy of multiparticle quantum batteries by analyzing the cases of three-particle and five-particle quantum batteries. Figure~\ref{BM4} plots the variation of the energy density (solid lines) and the average ergotropy (dashed lines) with interparticle interaction  $J$ at different temperatures and the dissipation rates of the battery $\Gamma_{B}$. Specifically, Figs.~\ref{BM4}(a)-\ref{BM4}(b) correspond to the case of the three-particle quantum battery, with dissipation rates of $\Gamma_{B}=0.05g$ and $\Gamma_{B}=0.5g$, respectively. Figures~\ref{BM4}(c)-\ref{BM4}(d) correspond to the case of the five-particle quantum battery. The curves in different colors represent different temperatures of the bosonic thermal reservoir, with the red curve corresponding to $T=0$ and the blue curve corresponding to $T=10$. The results indicate that, regardless of whether the quantum battery operates under a weak or strong dissipation regime, interparticle interactions tend to degrade the performance of the multiparticle quantum battery. Compared to the weak dissipation regime, strong dissipation significantly impairs the energy storage and extraction capabilities of the quantum battery. However, under a strong dissipation regime, a high-temperature environment can enhance both energy storage and extraction.
\begin{figure*}
\centering\scalebox{0.4}{\includegraphics{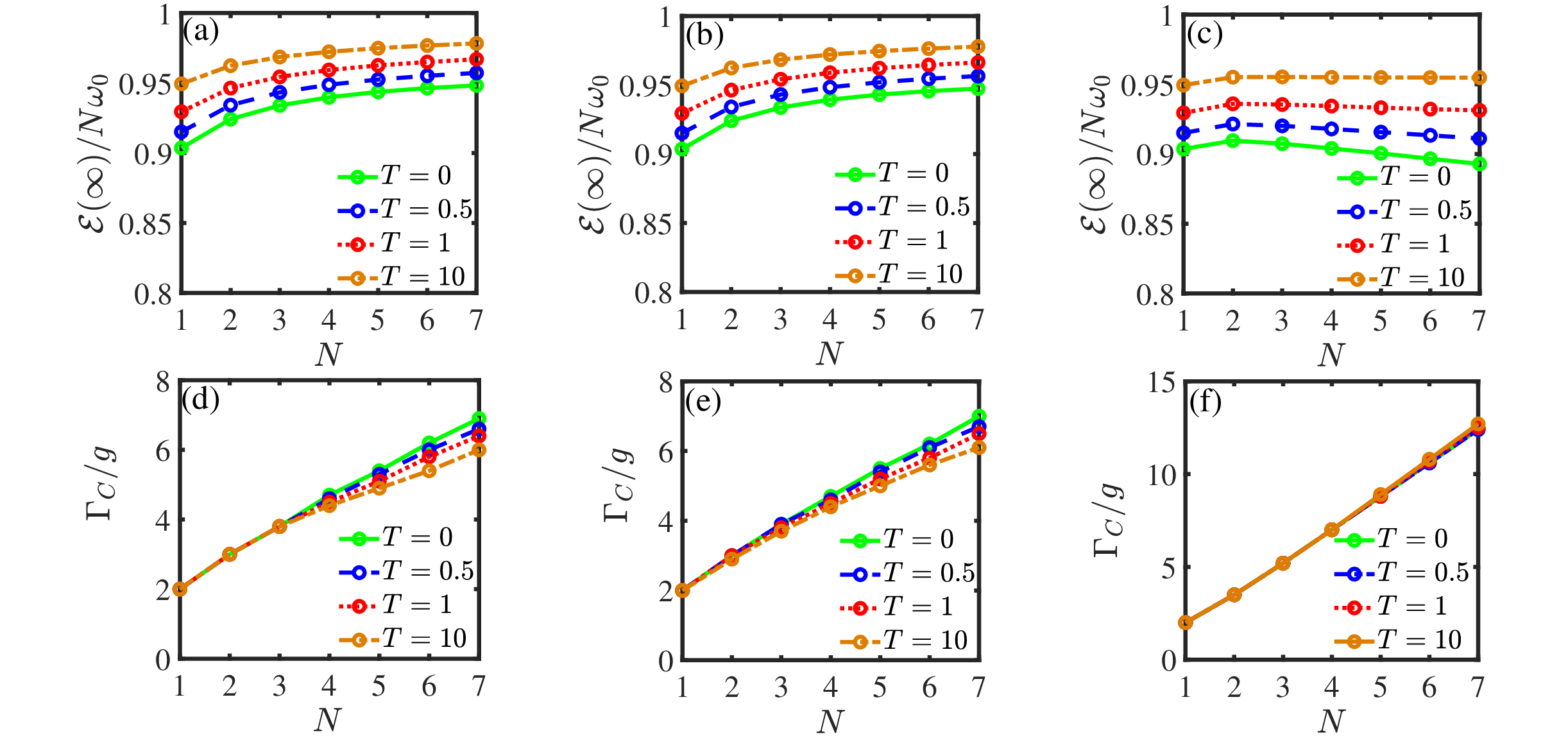}}
\caption{\label{FM2} The multiparticle quantum battery is embedded in a fermionic thermal reservoir. The average ergotropy of the multiparticle quantum battery is plotted as a function of the number of particles $N$, with all other parameters consistent with those in Fig.~\ref{BM1}.
 }
\end{figure*}
\begin{figure*}
\centering\scalebox{0.4}{\includegraphics{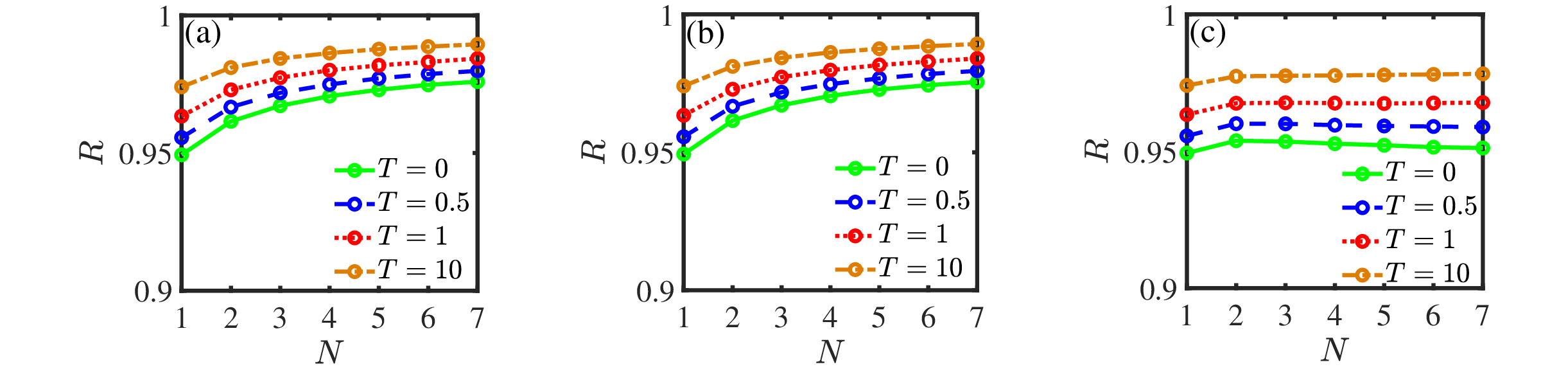}}
\caption{\label{FM3} Variation in the charging efficiency $R$ of the multiparticle quantum battery in the steady state with the number of particles $N$ in a fermionic thermal reservoir for (a) $J=0$, (b) $J=0.1g$,and (c) $J=g$. The other parameters are the same as those in Fig.~\ref{BM1}.
 }
\end{figure*}
\begin{figure}
\centering\scalebox{0.32}{\includegraphics{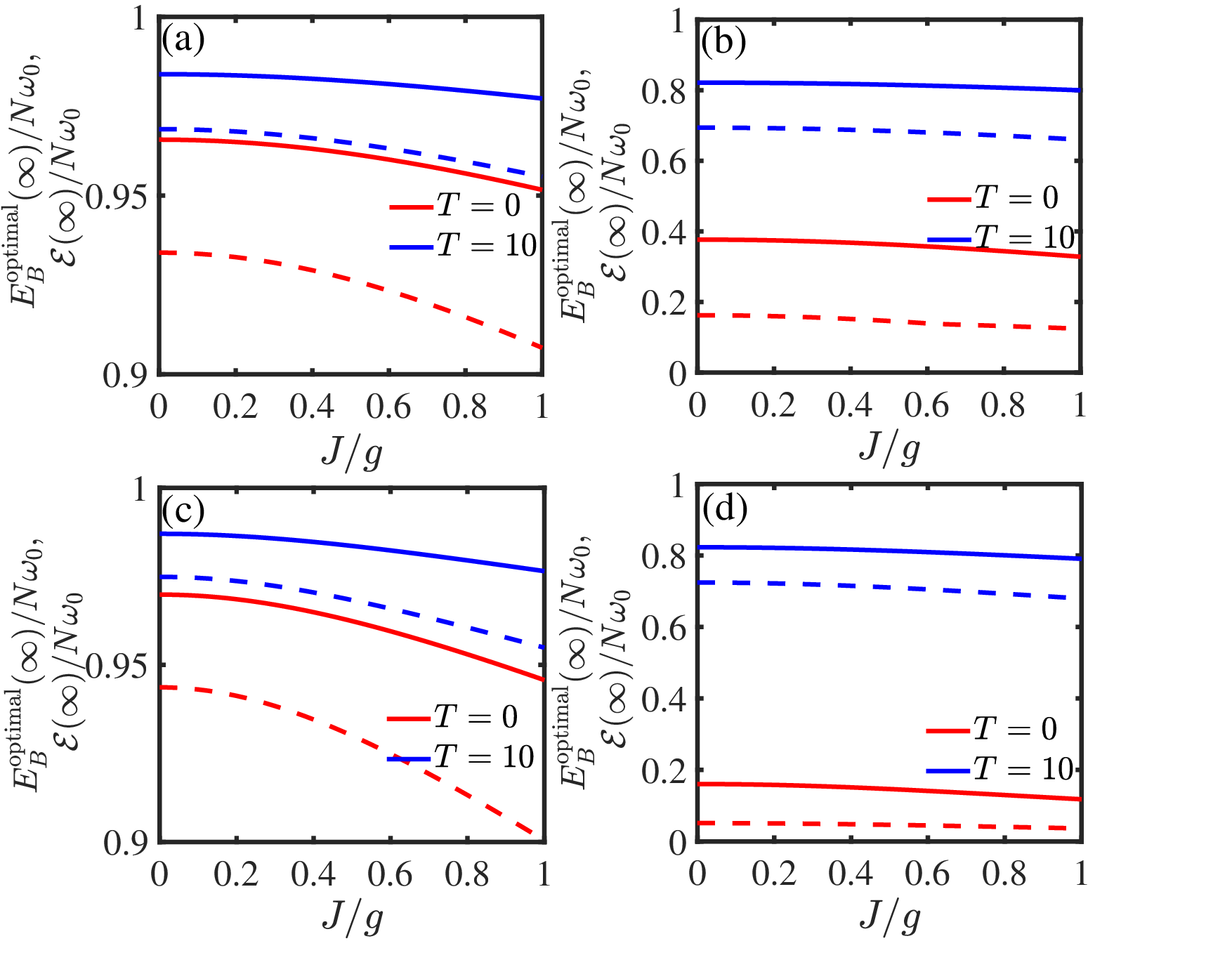}}
\caption{\label{FM4} Similar to Fig.~\ref{BM4}, variations in the energy density and the average ergotropy are shown for the (a) and (b) three-particle and (c) and (d) five-particle quantum batteries under (a) and (c) weak dissipation rate $\Gamma_{B}=0.05g$ and (b) and (d) strong dissipation rate $\Gamma_{B}=0.5g$, as a function of the interparticle coupling strength $J$ of the battery. The red curve corresponds to $T=0$ ($n_{f}=0$) and the blue curve corresponds to $T=10$ [$n_{f}=1/(e^{1/10}+1)\approx0.475$].
 }
\end{figure}
\subsection{Charging process of the multiparticle quantum battery in a fermionic thermal reservoir}
Next we investigate the charging process of the multiparticle quantum battery coupled to a fermionic thermal reservoir. In this scenario, $\gamma^{\downarrow}_{B}=\Gamma_{B}(1-n_{f})$ and $\gamma^{\uparrow}_{B}=\Gamma_{B}n_{f}$ in Eq.~(\ref{LB}).

First, we explore the effects of a finite-temperature fermionic thermal reservoir on the energy storage and maximum extractable energy (ergotropy) of the multiparticle quantum battery, under the weak dissipation condition ($\Gamma_{B}=0.05g$). Furthermore, we examine the effects of the interparticle interaction strength $J$ and the number of particles $N$ on the battery performance, aiming to identify a more efficient charging mechanism.

Figures~\ref{FM1}(a)-\ref{FM1}(c) illustrate the variation of energy density (i.e., the stored energy per particle) of the multiparticle quantum battery as a function of the number of particles $N$, under three different conditions, respectively: no coupling ($J=0$), weak coupling ($J=0.1g$), and strong coupling ($J=g$) between the internal particles of the battery. The optimal charger dissipation rates corresponding to Figs.~\ref{FM1}(a)-\ref{FM1}(c) are presented in Figs.~\ref{FM1}(d)-\ref{FM1}(f). The numerical results show that when the interaction strength $J$ between particles is relatively weak ($J=0$ or $0.1g$), increasing the number of battery particles enhances the energy density of the multiparticle quantum battery. Moreover, a higher temperature also leads to an increase in energy density. Similar to the single-particle battery case, temperature remains a favorable factor in the charging process. In the case of a relatively strong interaction strength $J$, it can be observed from Fig.~\ref{FM1} (c) that the energy density of the multiparticle quantum battery decreases as the number of particles $N$ increases under low-temperature conditions. As the temperature increases, the impact of the number of particles on energy storage becomes less pronounced. Furthermore, Figs.~\ref{FM1}(d)-\ref{FM1}(f) indicate that the optimal dissipation rate $\Gamma_{C}$ of the charger required to maximize energy storage increases when the number of battery particles $N$ increases. 

Figure~\ref{FM2} illustrates the impact of the reservoir temperature, particle number $N$, and the interparticle interaction $J$ on the average ergotropy of the multiparticle quantum battery in a finite-temperature fermionic thermal reservoir. The results indicate that the behavior of the ergotropy closely resembles that of the energy storage in the battery. Figure~\ref{FM3} reveals the influences of the temperature and the particle number $N$ on the charging efficiency $R$ of the multiparticle quantum battery. First, it can be observed that an increase in temperature enhances the charging efficiency $R$ of the multiparticle quantum battery. Second, when the interparticle coupling strength is weak ($J=0$ or $0.1g$), $R$ increases as the particle number $N$ increases. However, when the interparticle coupling strength $J$ is strong ($J=g$), an increase in $N$ leads to a reduction in the charging efficiency $R$ at low temperatures. In contrast, at high temperatures, the effect of increasing particle number on the charging efficiency becomes less pronounced. Based on the above research findings, we observe that when the dissipation rate of the multiparticle quantum battery in a fermionic thermal reservoir is weak, higher temperatures and weaker interparticle coupling enhance the performance of the quantum battery. The case of a strongly dissipative quantum battery is discussed in the Appendix~\ref {D}.

Subsequently, we continue to examine the impact of interparticle interaction $J$ on both the energy density and average ergotropy of the multiparticle quantum battery, using three-particle and five-particle quantum batteries as representative cases, as shown in Fig.~\ref{FM4}. Figures~\ref{FM4}(a)-\ref{FM4}(b) display the numerical results for the three-particle quantum battery with dissipation rates $\Gamma_{B}=0.05g$ and $\Gamma_{B}=0.5g$, respectively, while Figs.~\ref{FM4}(c)-\ref{FM4}(d) show the results for the corresponding five-particle quantum battery under identical dissipation conditions. The red and blue curves correspond to the fermionic thermal reservoir temperatures of the $T=0$ and $10$, respectively. The solid lines represent the energy density of the battery, while the dashed lines represent the average ergotropy. Numerical analysis reveals that when the multiparticle quantum battery is coupled to a fermionic thermal reservoir, interparticle interactions $J$ do not enhance its performance, regardless of whether the battery operates in a weak or strong dissipation regime. However, compared to low-temperature conditions, a high-temperature environment significantly improves both the energy storage capacity and the extractable energy of the battery.

Based on the above discussion, regardless of whether the multiparticle quantum battery is placed in a bosonic or fermionic thermal reservoir, both the energy density and the average ergotropy of the multiparticle quantum battery decrease as the interparticle interaction strength $J$ increases. The underlying physical mechanism behind this behavior lies in the interaction-induced splitting of the battery's energy eigenstates. This spectral reconstruction shifts the system away from the resonance, thereby suppressing energy transfer from the charger to the quantum battery. Consequently, interparticle interactions within the battery emerge as a significant obstacle in designing and implementing efficient multiparticle quantum batteries. Additionally, a high-temperature fermionic thermal reservoir provides a stable foundation for building multiparticle quantum batteries.

\section{Summary}\label{V}
In this study, we investigated charging models for single-particle and multiparticle quantum batteries based on homodyne quantum feedback control. In the single-particle model, both the charger and the battery were modeled as two-level atomic systems. Quantum feedback control was applied to the charger during the charging process, enabling energy transfer through the coupling between the charger and the battery. Subsequently, we examined the effects of bosonic and fermionic thermal reservoirs on the performance of the single-particle quantum battery. The findings reveal the existence of optimal charging parameters for the system. Compared to bosonic thermal reservoirs, fermionic reservoirs at finite temperature facilitate energy storage and extraction more efficiently. Moreover, the battery charging efficiency $R$ improves with increasing temperature. In the multiparticle quantum battery model, the charger is still modeled as a two-level system, while the quantum battery consists of an ensemble of two-level systems. We further analyzed the charging process when the multiparticle quantum battery is collectively placed in either a bosonic or fermionic thermal reservoir. Additionally, we investigated the effects of the reservoir temperature $T$, the interparticle coupling strength $J$, and the number of particles $N$ on the performance of the multiparticle quantum battery. The results indicate that when the multiparticle quantum battery is placed in a bosonic thermal reservoir, increasing the temperature reduces both the energy density and the average ergotropy, consistent with the behavior observed in the single-particle quantum battery. Furthermore, expanding the size of the quantum battery in a bosonic thermal reservoir requires satisfying three conditions: a lower thermal reservoir temperature, weak inter-atomic interactions within the battery, and a low dissipation rate of the quantum battery. For the multiparticle quantum battery in a fermionic thermal reservoir, high temperatures still exhibit corresponding advantages in terms of the battery's energy density and average ergotropy. Simultaneously, the conditions for constructing a multiparticle quantum battery are significantly relaxed. Even under strong interparticle interactions and higher battery dissipation rates, the multiparticle quantum battery still maintains relatively high-performance levels in both energy storage and energy extraction under high-temperature conditions.

\section*{Acknowledgements}
This work was supported by the National Natural Science Foundation of China under Grant No. 12174048 and by Shenyang Ligong University through high-level talent research support funds under Grant No. 1010147001309.
\section*{Data availability}
The data that support the findings of this article are openly available~\cite{statement}.
\appendix
\begin{widetext}
\section{Steady-state solution of the system when the quantum battery is in a bosonic and fermionic thermal reservoir}\label{A}
To solve Eq.~(\ref{master}), we expand all the operators appearing in it using a global basis set for the two qubits. We choose $|1\rangle=|ee\rangle$, $|2\rangle=|eg\rangle$, $|3\rangle=|ge\rangle$, and $|4\rangle=|gg\rangle$. Accordingly, the density operator $\rho(t)$ of the system at an arbitrary time $t$ can be written as $\rho(t)=\sum_{i,j=1}^{4}\rho_{ij}(t)|i\rangle \langle j|$, in matrix form,
\begin{equation}\label{JUZHEN}
\rho(t)=\left(
\begin{array}{cccc}
\rho_{11}(t)&\rho_{12}(t)&\rho_{13}(t)&\rho_{14}(t)\\
 \rho_{21}(t)&\rho_{22}(t)&\rho_{23}(t)&\rho_{24}(t)\\
 \rho_{31}(t)&\rho_{32}(t)&\rho_{33}(t)&\rho_{34}(t)\\
 \rho_{41}(t)&\rho_{42}(t)&\rho_{43}(t)&\rho_{44}(t)
\end{array}
\right),
\end{equation}
where $\rho_{ij}$ are the corresponding expansion coefficients. When the quantum battery is in contact with a bosonic thermal reservoir, the differential equations are obtained by solving the corresponding master equation
\begin{equation}
\dot{\rho}_{11}(t)=2f\rho_{11}(t)-(\Gamma_{C}+\Gamma_{B}+n_{b}\Gamma_{B})\rho_{11}(t)+n_{b}\Gamma_{B}\rho_{22}(t)+\frac{f^{2}}{\Gamma_{C}\eta}[\rho_{33}(t)-\rho_{11}(t)],
\end{equation}
\begin{equation}
\dot{\rho}_{22}(t)=2f\rho_{22}(t)-ig[\rho_{32}(t)-\rho_{23}(t)]-(\Gamma_{C}+n_{b}\Gamma_{B})\rho_{22}(t)+(1+n_{b})\Gamma_{B}\rho_{11}(t)+\frac{f^{2}}{\Gamma_{C}\eta}[\rho_{44}(t)-\rho_{22}(t)],
\end{equation}
\begin{equation}
\dot{\rho}_{33}(t)=-2f\rho_{11}(t)-ig[\rho_{23}(t)-\rho_{32}(t)]+\Gamma_{C}\rho_{11}(t)-(1+n_{b})\Gamma_{B}\rho_{33}(t)+n_{b}\Gamma_{B}\rho_{44}(t)+\frac{f^{2}}{\Gamma_{C}\eta}[\rho_{11}(t)-\rho_{33}(t)],
\end{equation}
\begin{equation}
\dot{\rho}_{44}(t)=-2f\rho_{22}(t)+\Gamma_{C}\rho_{22}(t)+(1+n_{b})\Gamma_{B}\rho_{33}(t)-n_{b}\Gamma_{B}\rho_{44}(t)+\frac{f^{2}}{\Gamma_{C}\eta}[\rho_{22}(t)-\rho_{44}(t)],
\end{equation}
\begin{equation}
\dot{\rho}_{12}(t)=\dot{\rho}_{21}^{*}(t)=2f\rho_{12}(t)+ig\rho_{13}(t)-\left[\Gamma_{C}+\left(n_{b}+\frac{1}{2}\right)\Gamma_{B}\right]\rho_{12}(t)+\frac{f^{2}}{\Gamma_{C}\eta}\left[\rho_{34}(t)-\rho_{12}(t)\right],
\end{equation}
\begin{equation}
\dot{\rho}_{13}(t)=\dot{\rho}_{31}^{*}(t)=ig\rho_{12}(t)-\left[\frac{\Gamma_{C}}{2}+(1+n_{b})\Gamma_{B}\right]\rho_{13}(t)+n_{b}\Gamma_{B}\rho_{24}(t)+\left(f-\frac{f^{2}}{\Gamma_{C}\eta}\right)[\rho_{31}(t)+\rho_{13}(t)],
\end{equation}
\begin{eqnarray}
\dot{\rho}_{14}(t)=\dot{\rho}_{41}^{*}(t)=-\left(\frac{\Gamma_{C}+\Gamma_{B}}{2}+n_{b}\Gamma_{B}\right)\rho_{14}(t)+\left(f-\frac{f^{2}}{\Gamma_{C}\eta}\right)[\rho_{32}(t)+\rho_{14}(t)],
\end{eqnarray}
\begin{equation}
\dot{\rho}_{23}(t)=\dot{\rho}_{32}^{*}(t)=-ig[\rho_{33}(t)-\rho_{22}(t)]-\left(\frac{\Gamma_{C}+\Gamma_{B}}{2}+n_{b}\Gamma_{B}\right)\rho_{23}(t)+\left(f-\frac{f^{2}}{\Gamma_{C}\eta}\right)[\rho_{23}(t)+\rho_{41}(t)],
\end{equation}
\begin{equation}
\dot{\rho}_{24}(t)=\dot{\rho}_{42}^{*}(t)=-ig\rho_{34}(t)-\left(\frac{\Gamma_{C}}{2}+n_{b}\Gamma_{B}\right)\rho_{24}(t)+(1+n_{b})\Gamma_{B}\rho_{13}(t)+\left(f-\frac{f^{2}}{\Gamma_{C}\eta}\right)[\rho_{24}(t)+\rho_{42}(t)],
\end{equation}
\begin{equation}
\dot{\rho}_{34}(t)=\dot{\rho}_{43}^{*}(t)=(\Gamma_{C}-2f)\rho_{12}(t)-ig\rho_{24}(t)-\left[\Gamma_{B}\left(\frac{1}{2}+n_{b}\right)\right]\rho_{34}(t)+\frac{f^{2}}{\Gamma_{C}\eta}[\rho_{12}(t)-\rho_{34}(t)].
\end{equation}
Then setting $\dot{\rho}(t)=0$, we obtain the steady-state solutions
\begin{equation}\label{ww}
\rho_{11}^{\infty}=\frac{4g^{2}Q_{b}^{2}+n_{b}\delta^{2}W_{b}}{(1+2n_{b})W_{b}S+4g^{2}[2Q_{b}+(\Gamma_{C}+\Gamma_{B}-2\Gamma_{C}\delta)\eta]^{2}},\quad \rho_{22}^{\infty}=\frac{4g^{2}Q_{b}\Gamma_{C}(S-\delta^{2})+(1+n_{b})(\delta^{2}W_{b}+4g^{2}Q_{b}\Gamma_{B}\eta)}{(1+2n_{b})W_{b}S+4g^{2}[2Q_{b}+(\Gamma_{C}+\Gamma_{B}-2\Gamma_{C}\delta)\eta]^{2}},
\end{equation}

\begin{equation}
\rho_{33}^{\infty}=\frac{4g^{2}Q_{b}[(1+n_{b})\Gamma_{B}\eta+\Gamma_{C}(S-\delta^{2})]+n_{b}W_{b}(S-\delta^{2})}{(1+2n_{b})W_{b}S+4g^{2}[2Q_{b}+(\Gamma_{C}+\Gamma_{B}-2\Gamma_{C}\delta)\eta]^{2}},\quad
\end{equation}
\begin{equation}
\rho_{44}^{\infty}=\frac{4g^{2}\Gamma_{C}(S-\delta^{2})+(1+n_{b})[W_{b}(S-\delta^{2})+4g^{2}\Gamma_{B}\eta]}{(1+2n_{b})W_{b}S+4g^{2}[2Q_{b}+(\Gamma_{C}+\Gamma_{B}-2\Gamma_{C}\delta)\eta]^{2}},
\end{equation}
\begin{equation}
\rho_{14}^{\infty}=\rho_{41}^{\infty*}=\frac{4ig\delta(\delta-\eta)\Gamma_{C}^{2}\Gamma_{B}[\delta^{2}+n_{b}(2\delta-1)\eta]}{(1+2n_{b})W_{b}S+4g^{2}[2Q_{b}+(\Gamma_{C}+\Gamma_{B}-2\Gamma_{C}\delta)\eta]^{2}},
\end{equation}
\begin{equation}
\rho_{23}^{\infty}=\rho_{32}^{\infty*}=\frac{2ig\Gamma_{C}\Gamma_{B}[\delta^{2}+n_{b}(2\delta-1)\eta]\left\{2Q_{b}+[\Gamma_{C}(1-2\delta)+\Gamma_{B}]\eta\right\}}{(1+2n_{b})W_{b}S+4g^{2}[2Q_{b}+(\Gamma_{C}+\Gamma_{B}-2\Gamma_{C}\delta)\eta]^{2}}.
\end{equation}
We set $\delta=f/\Gamma_{C}$, $S=2\delta^{2}+\eta-2\delta\eta$, $Q_{b}=\Gamma_{C}\delta^{2}+n_{b}\Gamma_{B}\eta$ and $W_{b}=\Gamma_{C}\Gamma_{B}(\Gamma_{C}+\Gamma_{B}+2n_{b}\Gamma_{B})[4Q_{b}+(\Gamma_{C}+\Gamma_{B}-4\Gamma_{C}\delta-2n_{b}\Gamma_{B})\eta]$ for simplicity. The other steady-state expansion coefficients are all zero, that is $\rho_{12}^{\infty}=\rho_{21}^{\infty}=\rho_{13}^{\infty}=\rho_{31}^{\infty}=\rho_{24}^{\infty}=\rho_{42}^{\infty}=\rho_{34}^{\infty}=\rho_{43}^{\infty}=0$. Combining Eqs.~(\ref{energy}) and (\ref{ergotropy1}), we obtain the explicit expression for the stored energy and the ergotropy of the quantum battery as follows
\begin{equation}\label{A18}
E_{B}(t)=\omega_{0}[\rho_{11}(t)+\rho_{33}(t)],
\end{equation}
\begin{eqnarray}\label{A19}
\mathcal{E}(t)&=&\frac{\omega_{0}}{2}(\sqrt{4|\rho_{12}(t)+\rho_{34}(t)|^{2}+[2(\rho_{11}(t)+\rho_{33}(t))-1]^{2}}+\{2[\rho_{11}(t)+\rho_{33}(t)]-1\}.
\end{eqnarray}
According to Eqs.~(\ref{A18}) and (\ref{A19}), the energy storage and the ergotropy of the quantum battery in the steady state can be determined. The expression for the ergotropy of the battery is more complex and will not be provided here, while the expression for energy storage is
\begin{equation}
E_{B}{(\infty)}=\omega_{0}\frac{n_{b}W_{b}S+4g^{2}Q_{b}[2Q_{b}+\Gamma_{C}(1-2\delta)\eta+\Gamma_{B}\eta]}{(1+2n_{b})W_{b}S+4g^{2}[2Q_{b}+\Gamma_{C}(1-2\delta)\eta+\Gamma_{B}\eta]^{2}}.
\end{equation}
When the quantum battery is in contact with a fermionic thermal reservoir, the differential equations are similarly obtained by solving the corresponding master equation. By setting $\dot{\rho}(t)=0$, the steady-state solution is obtained as

\begin{equation}
\rho_{11}^{\infty}=\frac{4g^{2}Q_{f}^{2}+n_{f}\delta^{2}W_{f}}{4g^2(\Gamma_{B}\eta+\Gamma_{C}S)^{2}+W_{f}S},\qquad
\rho_{22}^{\infty}=\frac{(1-n_{f})\delta^{2}W_{f}+4g^{2}Q_{f}[(1-n_{f})\Gamma_{B}\eta+\Gamma_{C}(S-\delta^{2})]}{4g^2(\Gamma_{B}\eta+\Gamma_{C}S)^{2}+W_{f}S},
\end{equation}
\begin{equation}
\rho_{33}^{\infty}=\frac{n_{f}(S-\delta^{2})W_{f}+4g^{2}Q_{f}[(1-n_{f})\Gamma_{B}\eta+\Gamma_{C}(S-\delta^{2})]}{4g^2(\Gamma_{B}\eta+\Gamma_{C}S)^{2}+W_{f}S},
\end{equation}
\begin{equation}
\rho_{44}^{\infty}=\frac{(1-n_{f})(S-\delta^{2})W_{f}+4g^{2}[(n_{f}-1)\Gamma_{B}\eta-\Gamma_{C}(S-\delta^{2})]^{2}}{4g^2(\Gamma_{B}\eta+\Gamma_{C}S)^{2}+W_{f}S},
\end{equation}
\begin{equation}
\rho_{14}^{\infty}=\rho_{41}^{\infty*}=-\frac{4ig\Gamma_{C}^{2}\delta\Gamma_{B}(\delta-\eta)[(2n_{f}-1)\delta^{2}+n_{f}\eta-2n_{f}\delta\eta]}{4g^2(\Gamma_{B}\eta+\Gamma_{C}S)^{2}+W_{f}S},
\end{equation}
\begin{equation}
\rho_{23}^{\infty}=\rho_{32}^{\infty*}=-\frac{2ig\Gamma_{C}\Gamma_{B}[(2n_{f}-1)\delta^{2}+n_{f}\eta(1-2\delta)](\Gamma_{B}\eta+\Gamma_{C}S)}{4g^2(\Gamma_{B}\eta+\Gamma_{C}S)^{2}+W_{f}S}.
\end{equation}
We set $\delta=f/\Gamma_{C}$ for simplicity. The other steady-state expansion coefficients are all zero, i.e. $\rho_{12}^{\infty}=\rho_{21}^{\infty}=\rho_{13}^{\infty}=\rho_{31}^{\infty}=\rho_{24}^{\infty}=\rho_{42}^{\infty}=\rho_{34}^{\infty}=\rho_{43}^{\infty}=0$.
According to Eqs.~(\ref{A18}) and (\ref{A19}), the stored energy and the ergotropy of the quantum battery in steady state can be determined. The expression for the ergotropy of the battery is more complex and will not be provided here, while the expression for energy storage is
\begin{equation}
E_{B}(\infty)=\omega_{0}\frac{4g^{2}Q_{f}(\Gamma_{B}\eta+\Gamma_{C}S)+n_{f}W_{f}S}{4g^2(\Gamma_{B}\eta+\Gamma_{C}S)^{2}+W_{f}S},
\end{equation}
where $\delta=f/\Gamma_{C}$, $S=2\delta^{2}+\eta-2\delta\eta$, $Q_{f}=\Gamma_{C}\delta^{2}+n_{f}\Gamma_{B}\eta$ and $W_{f}=\Gamma_{C}\Gamma_{B}(\Gamma_{C}+\Gamma_{B})[\Gamma_{B}\eta+\Gamma_{C}(2S-\eta)]$.
\end{widetext}

\begin{widetext}
\section{Comparative analysis between the present work and Ref.~ \cite{Song_2024}}\label{B}
In Ref.~ \cite{Song_2024}, the charger is modeled as a dissipative system embedded in either a bosonic or fermionic thermal reservoir, and powered by a driving laser field, while the quantum battery is isolated. For the case where the charger is embedded in a fermionic thermal reservoir with positive temperatures, we evaluate the quantum battery's performance and conduct a comparative analysis with our proposed scheme. When the system reaches a steady state, the stored energy $E_{B}(\infty)$ of the battery and the ergotropy $\mathcal{E}(\infty)$ of the battery are,
\begin{equation}\label{B1}
E_{B}(\infty)=\omega_{0}\frac{4F^{4}g^{2}+2A\Gamma_{C}^{2}+B\Gamma_{C}^{4}}{8F^{4}g^{2}+4[A+2g^{4}(1-2n_{f})]\Gamma_{C}^{2}+2[B+2g^{2}(1-2n_{f})]\Gamma_{C}^{4}},
\end{equation}
\begin{eqnarray}\label{B2}
\mathcal{E}(\infty)&=&\frac{\omega_{0}}{2}\left(-1+2\sqrt{\frac{g^{2}[B+g^{2}(1-4n_{f})](1-2n_{f})^{2}\Gamma_{C}^{4}(2g^{2}+\Gamma_{C}^{2})^{2}}{\{4F^{4}g^{2}+2[A+2g^{4}(1-2n_{f})]\Gamma_{C}^{2}+[B+2g^{2}(1-2n_{f})]\Gamma_{C}^{4}\}^{2}}}\right.\nonumber\\&&
\left.+\frac{4F^{4}g^{2}+2A\Gamma_{C}^{2}+B\Gamma_{C}^{4}}{4F^{4}g^{2}+2[A+2g^{4}(1-2n_{f})]\Gamma_{C}^{2}+[B+2g^{2}(1-2n_{f})]\Gamma_{C}^{4}}\right),
\end{eqnarray}
\end{widetext}

\begin{figure}
\centering\scalebox{0.36}{\includegraphics{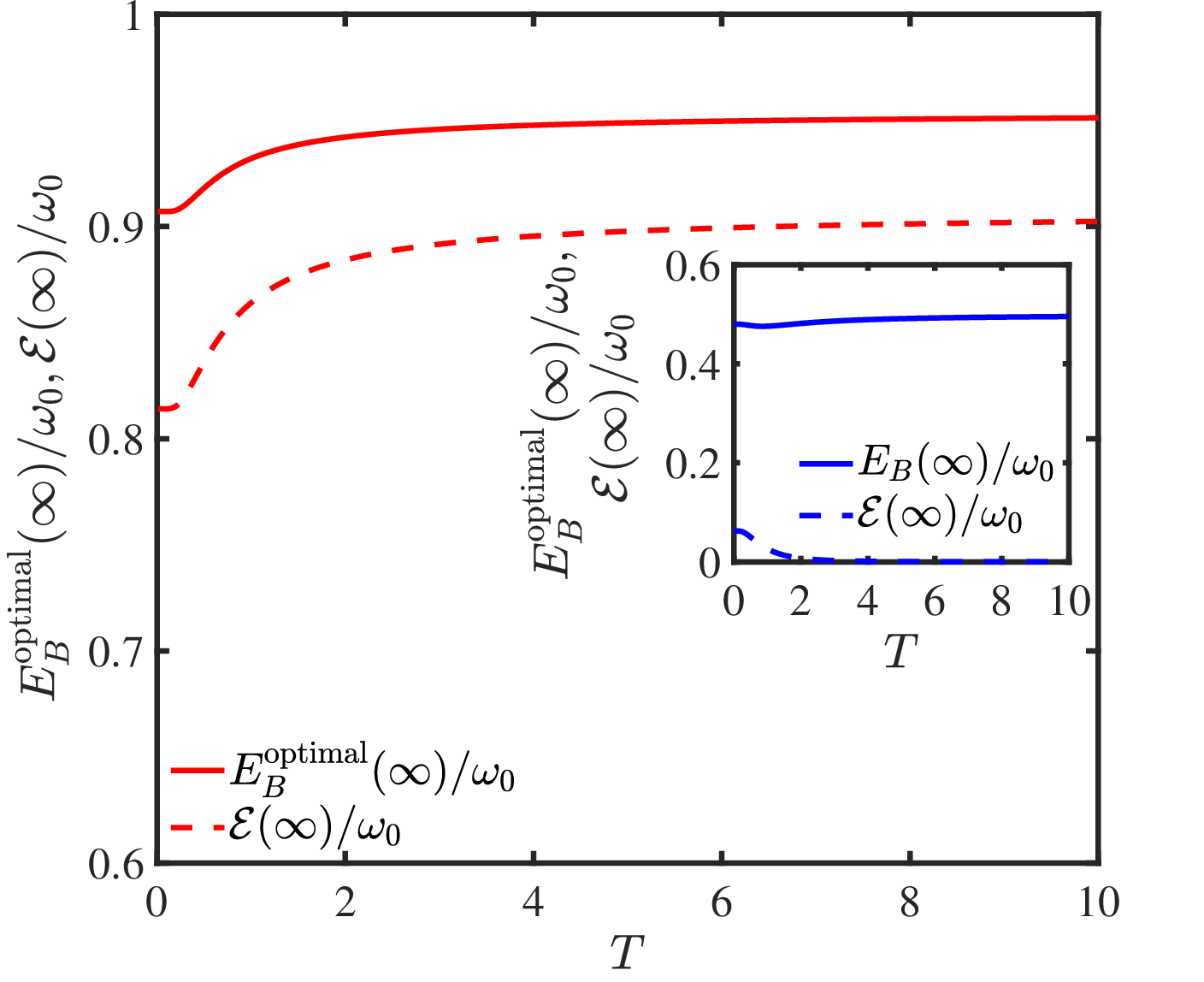}}
\caption{\label{f1} Steady-state stored energy (solid line) and ergotropy (dashed line) of the quantum battery plotted as a function of the fermionic thermal reservoir temperature under optimal charging parameters $\Gamma_{C}=2g$ and $\delta=f/\Gamma_{C}=1$. The coupling strength between the charger and the battery is set to $g=0.01\omega_{0}$. The red curve represents the results from our proposed scheme. The corresponding results from Ref.~\cite{Song_2024} are shown as the blue curve in the inset, with the parameter choices matching those in Fig.~5(a) of Ref.~\cite{Song_2024}, i.e.,$F=\Gamma_{C}=4g$.
 }
\end{figure}
\begin{figure*}
\centering\scalebox{0.4}{\includegraphics{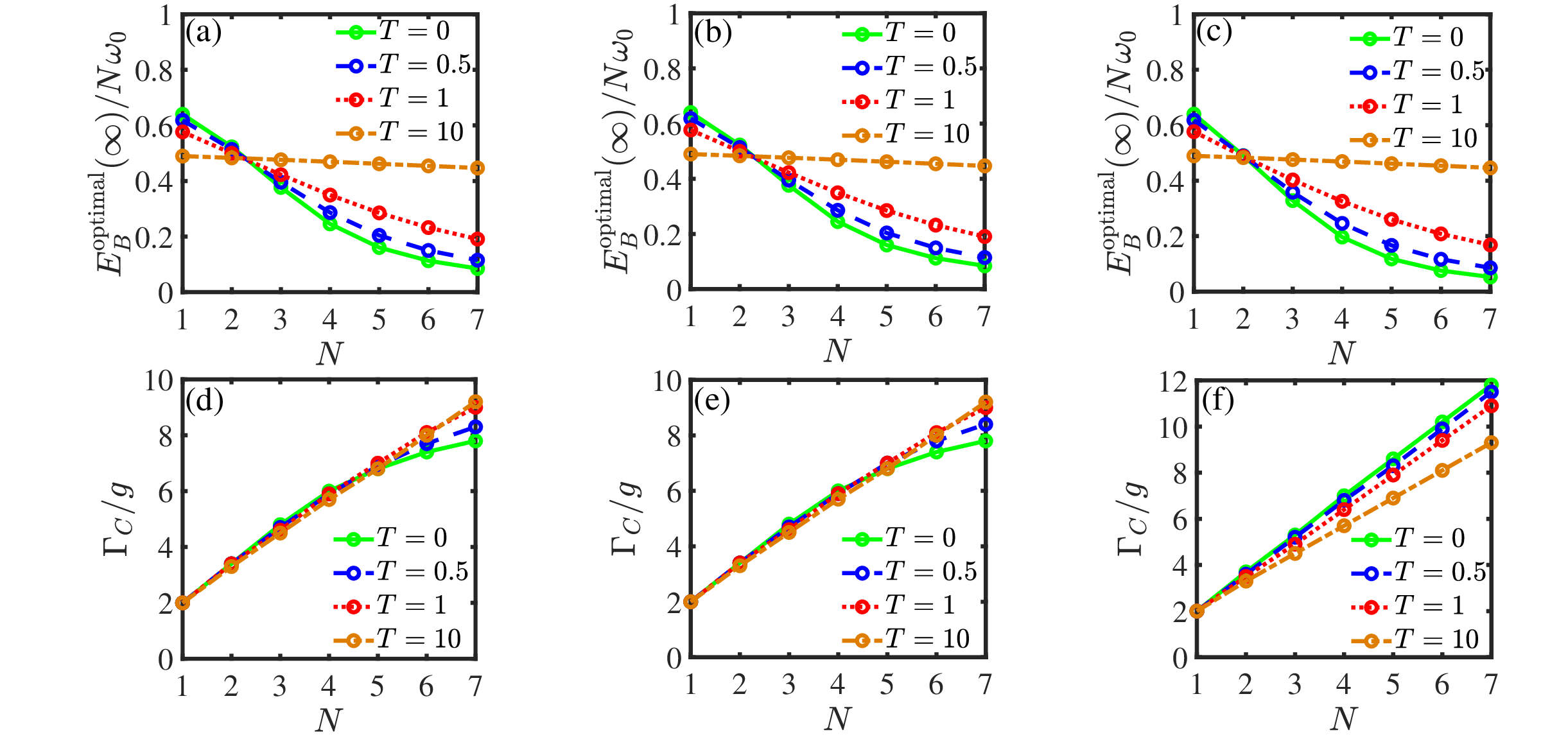}}
\caption{\label{BBM1} The multiparticle quantum battery is placed in a bosonic thermal reservoir, with a strong dissipation rate of $\Gamma_{B}=0.5g$. The optimal energy density of the battery is plotted as a function of the number of quantum battery particles, under different reservoir temperatures and internal interactions between particles within the battery. The remaining information is provided in Fig.~\ref{BM1}.
 }
\end{figure*}
\begin{figure*}
\centering\scalebox{0.4}{\includegraphics{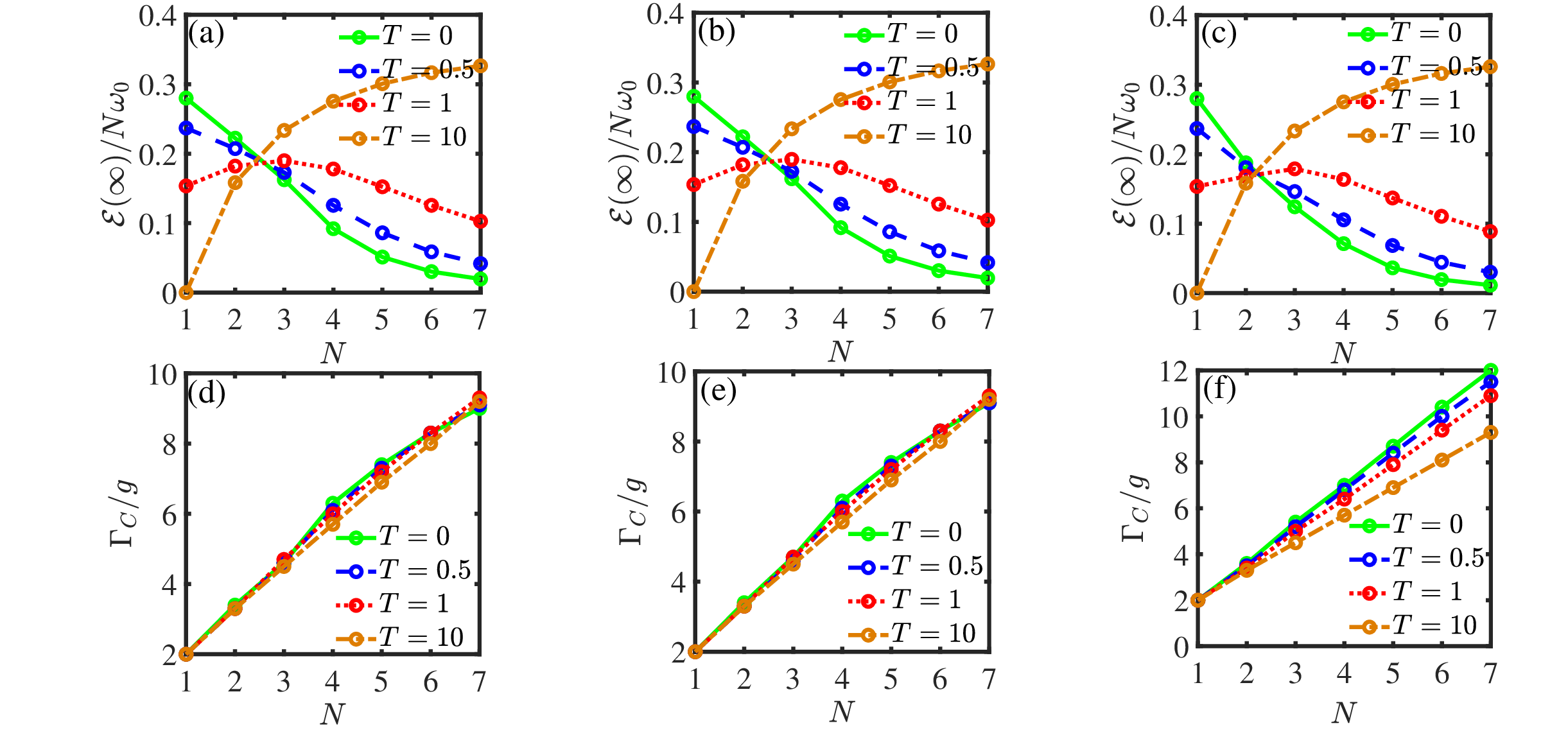}}
\caption{\label{BBM2} In a bosonic thermal reservoir, the average ergotropy of the quantum battery in the steady state is shown as a function of the particle number $N$, where $\Gamma_{B}=0.5g$. The remaining information is the same as in Fig.~\ref{BM2}.
 }
\end{figure*}
\begin{figure*}
\centering\scalebox{0.4}{\includegraphics{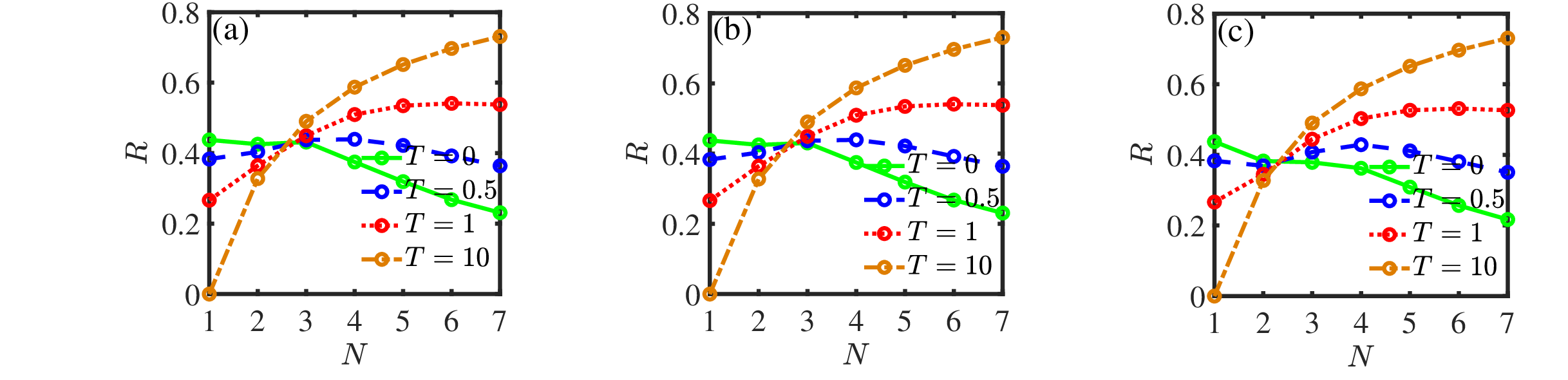}}
\caption{\label{BBM3} The multiparticle quantum battery is in a bosonic thermal reservoir. The charging efficiency $R$ of the battery is shown as a function of the number of battery particles $N$. The remaining information is the same as in Fig.~\ref{BM3}.
 }
\end{figure*}
respectively, where $A=F^{2}g^{2}[1-8(n_{f}-1)n_{f}]+F^{4}(1-2n_{f})^2+4g^{4}n_{f}$ and $B=F^{2}(1-2n_{f})^{2}+4g^{2}n_{f}$. Here $F$ is the amplitude of the driving field, $\Gamma_{C}$ is the dissipation rate of the charger, $g$ is the coupling strength between the charger and the battery, and $n_{f}$ is the average excitation number of the fermionic thermal reservoir.
\begin{figure*}
\centering\scalebox{0.4}{\includegraphics{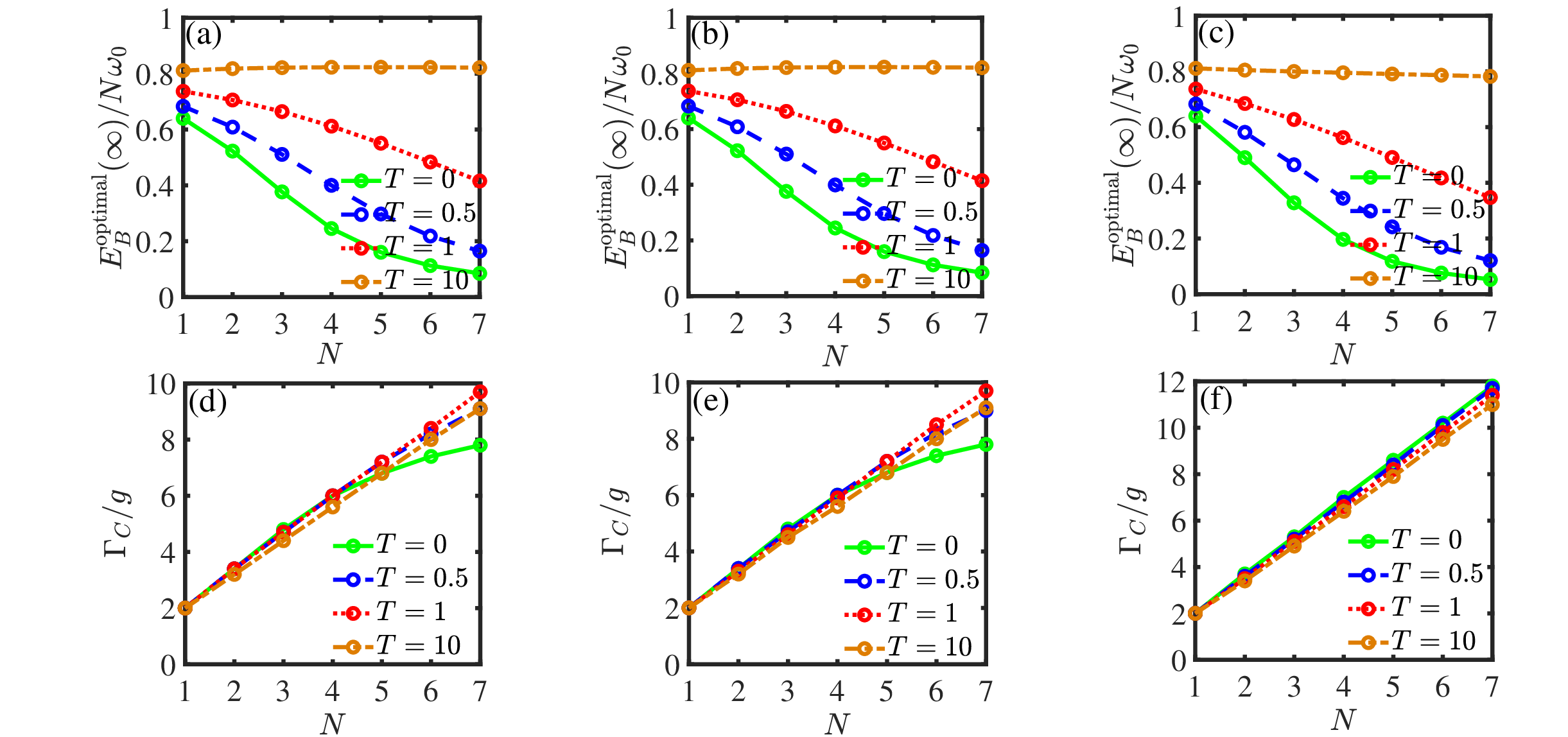}}
\caption{\label{FFM1} The multiparticle quantum battery is placed in a fermionic thermal reservoir. The optimal energy density of multiparticle quantum batteries is shown as a function of the particle number $N$. The only difference from Fig.~\ref{FM1} is that the dissipation rate of the battery is set to $\Gamma_{B}=0.5g$.
 }
\end{figure*}
To better compare the differences between our scheme and that in the Ref.~\cite{Song_2024}, Fig.~\ref{f1} plots the steady-state stored energy of the quantum battery (red solid line) and its ergotropy (red dashed line) under our charging scheme in a fermionic thermal reservoir as a function of temperature. The corresponding results from Ref.~\cite{Song_2024} are displayed in the inset, with the parameter choices matching those in Fig.~5(a) of Ref.~\cite{Song_2024}, i.e., $F=4g$ and $\Gamma_{C}=4g$. Here, the solid line represents the steady-state energy storage of the battery, and the dashed line corresponds to the maximum extractable energy (the ergotropy) of the battery. The results indicate that, in our scheme, both the energy storage and the maximum extractable energy of the battery increase with the temperature rise. In contrast, the results in Ref.~\cite{Song_2024} show that, as the temperature increases, the maximum stored energy $E_{B}^{\rm optimal}(\infty)$ of the quantum battery tends to approach $0.5\omega_{0}$, and the maximum extractable energy $\mathcal{E}(\infty)$ gradually decreases to zero.
\section{A strongly dissipative multiparticle quantum battery in a bosonic thermal reservoir}\label{C}
\begin{figure*}
\centering\scalebox{0.4}{\includegraphics{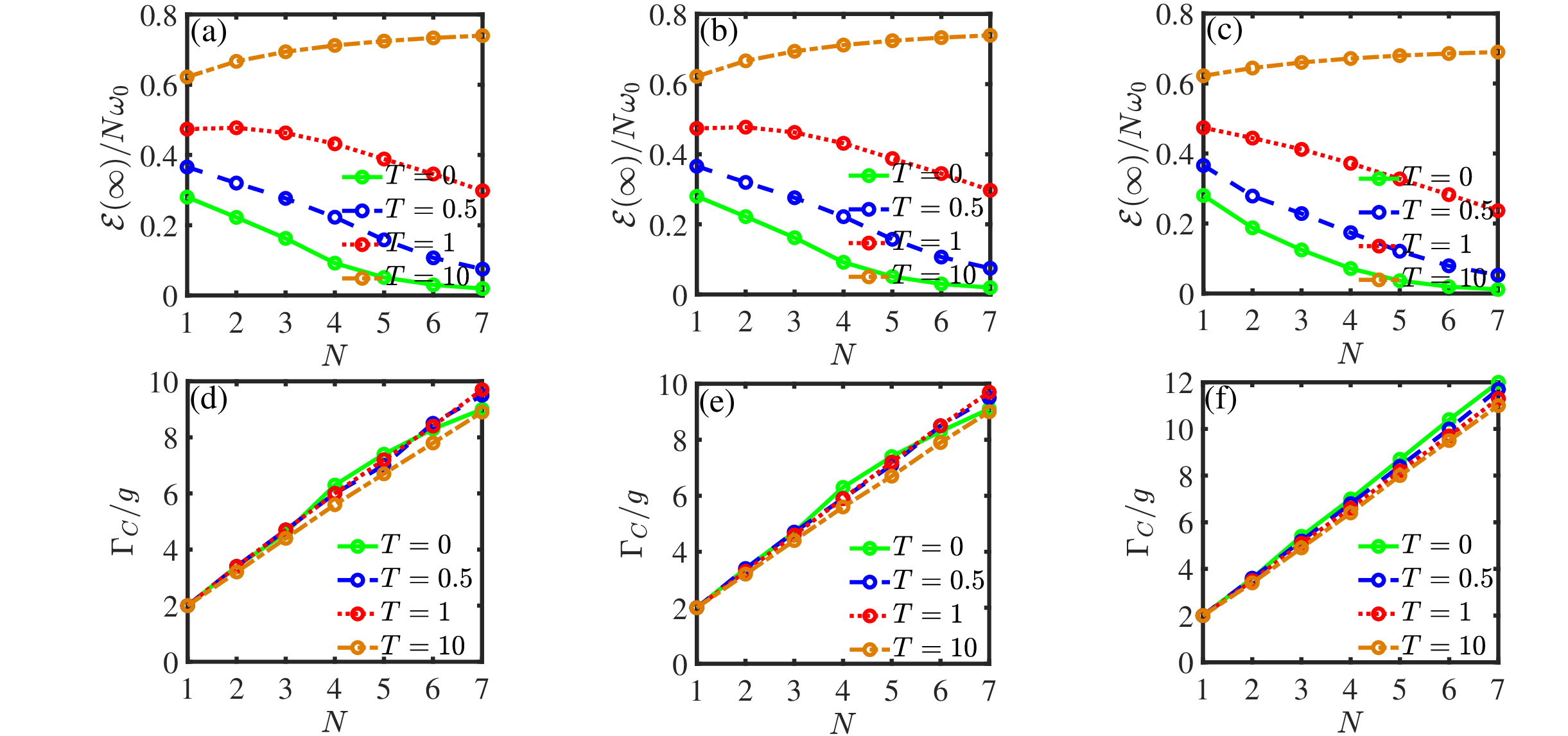}}
\caption{\label{FFM2} In a fermionic thermal reservoir, the average ergotropy of the multiparticle quantum battery in the steady state is shown as a function of the particle number $N$. Except for $\Gamma_{B}=0.5g$, all other parameters are the same as in Fig.~\ref{FM2}.
 }
\end{figure*}
\begin{figure*}
\centering\scalebox{0.4}{\includegraphics{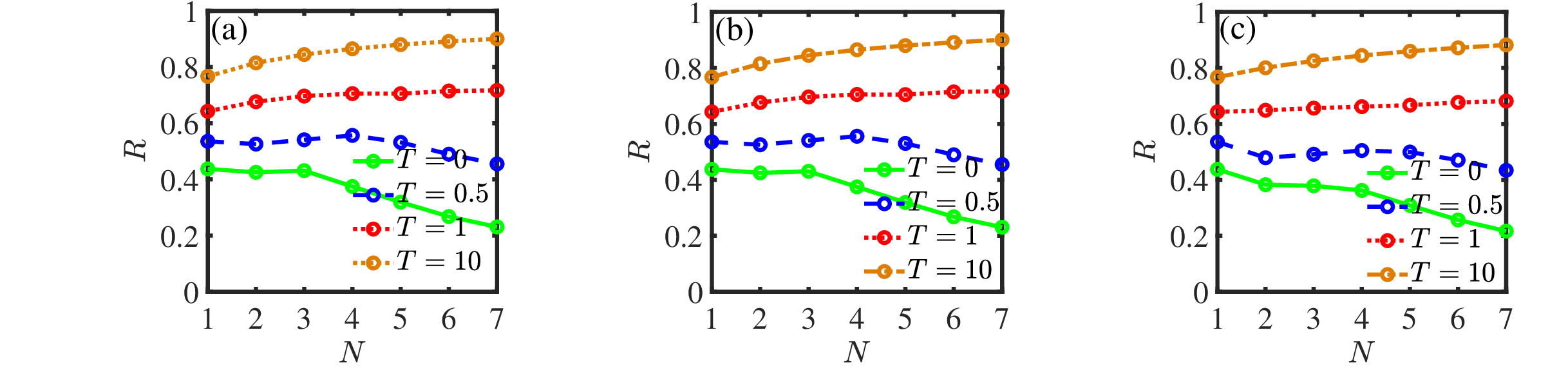}}
\caption{\label{FFM3} The multiparticle quantum battery is coupled to a fermionic thermal reservoir. The charging efficiency $R$ of the battery is shown as a function of the number of battery particles $N$, where $\Gamma_{B}=0.5g$, and the other parameters are the same as those in Fig.~\ref{FM3}.
 }
\end{figure*}
In this Appendix, we investigate the dependence of the energy density, the average ergotropy, and the charging efficiency $R$ of the multiparticle quantum battery on three key parameters, i.e., the thermal reservoir temperature, the number of particles $N$, and the interatomic interaction strength $J$, under strong dissipation conditions ($\Gamma_{B}=0.5g$) in a bosonic thermal reservoir. Figure~\ref{BBM1} shows how the energy density of the multiparticle quantum battery varies with the particle number $N$ under different temperatures and interatomic interactions within the battery. Figures~\ref{BBM1}(a)-\ref{BBM1}(c) correspond to $J=0$, $0.1g$, and $g$, respectively, and the different curves represent different thermal reservoir temperatures. Figures~\ref{BBM1}(d)-\ref{BBM1}(f) correspond to the optimal dissipative rates of the charger required for Figs.~\ref{BBM1}(a)-\ref{BBM1}(c). Unlike the weak dissipation regime depicted in Fig.~\ref{BM1}, the strong dissipation regime exhibits a universal decrease in energy density with increasing particle number $N$, independent of both the thermal reservoir temperature (ranging from low to high) and the strength of interparticle coupling within the battery (whether weak or strong). Consequently, constructing multiparticle quantum batteries in a strongly dissipative bosonic thermal reservoir is not advisable. However, it should be noted that for the number of particles $N>2$, a high-temperature environment can alleviate the reduction in energy density induced by increasing $N$. Figure~\ref{BBM2} shows correspondingly the behavior of the average ergotropy of the multiparticle quantum battery. The result shows that in a high-temperature environment, an increase in the number of particles $N$ is beneficial for energy extraction. Meanwhile, as shown in Fig.~\ref{BBM3}, the charging efficiency
$R$ of the battery increases with the number of particles under high-temperature conditions. In contrast, under low-temperature conditions, $R$ decreases as the number of particles $N$ increases.

\section{A strongly dissipative multiparticle quantum battery in a fermionic thermal reservoir.}\label{D}
Figure~\ref{FFM1} shows the energy density of the battery as a function of the number of particles $N$ for the multiparticle quantum battery operating in a strongly dissipative fermionic thermal reservoir ($\Gamma_{B}=0.5g$). Figures~\ref{FFM1}(a)-\ref{FFM1}(c) correspond to $J=0$, $0.1g$, and $g$, respectively, with different curves representing various temperatures of the thermal reservoir. Figures~\ref{FFM1}(d)-\ref{FFM1}(f) show the corresponding optimal dissipative rates of the charger required for the cases in Figs.~\ref{FFM1}(a)-\ref{FFM1}(c), respectively. The results indicate that in a strong fermionic thermal reservoir, an increase in temperature enhances the energy density of the quantum battery. This means that higher temperatures facilitate energy storage in multiparticle quantum batteries. However, increasing the number of particles in the battery under low temperature conditions will have a negative impact on the energy density, regardless of whether the interatomic interaction strength is strong or weak. Conversely, under high-temperature conditions, as the number of battery particles increases, the battery will maintain a state with higher energy density, and the interatomic interaction strength $J$ has a relatively minor influence on its energy density. As shown in Fig.~\ref{FFM2}, the behavior of the average ergotropy closely resembles that of energy storage. Furthermore, Fig.~\ref{FFM3} shows that the charging efficiency $R$ of the battery increases with the number of particles under high-temperature conditions, whereas under low-temperature conditions, $R$ decreases as $N$ increases. 

Overall, regardless of whether the multiparticle quantum battery operates under a low or high dissipation rate, its performance in a high-temperature fermionic thermal reservoir is significantly superior to that of a counterpart constructed in a bosonic thermal reservoir.

\bibliography{yao.bib}

\end{document}